\documentclass[preprint,10pt,authoryear]{elsarticle}

\usepackage{amssymb}
\usepackage{amsmath}
\usepackage{tikzfill}
\usepackage{soul} 
\usepackage[margin=1in]{geometry}
\usepackage{mathtools}
\usepackage{amsthm}
\usepackage{amssymb}
\usepackage{tikz}
\usepackage{pgfplots}
\pgfplotsset{compat=1.18} 
\usepackage{caption}
\usepackage{subcaption}
\usepackage{comment}
\usepackage{algorithm}
\usepgfplotslibrary{groupplots}
\usepackage[]{algpseudocode}
\usepgfplotslibrary{fillbetween}
\usepackage{tabularx} 
\usetikzlibrary{decorations.markings}
\definecolor{dark green}{rgb}{0.0, 0.4, 0.0}

\newtheorem{theorem}{Theorem}
\newtheorem{lemma}{Lemma}
\theoremstyle{definition}
\newtheorem{definition}{Definition}
\newtheorem{corollary}{Corollary}[lemma] 
\theoremstyle{remark}

\newcommand{\ee}{\mathbf{e}}
\newcommand{\xx}{\mathbf{x}}
\newcommand{\yy}{\mathbf{y}}

\newcommand{\bb}{\mathbf{b}}
\newcommand{\nn}{\mathbf{n}}
\newcommand{\uu}{\mathbf{u}}
\newcommand{\UU}{\mathbf{U}}
\newcommand{\vv}{\mathbf{v}}
\newcommand{\ww}{\mathbf{w}}

\newcommand{\HH}{\mathbf{H}}

\newcommand{\LL}{\mathcal{L}}

\renewcommand{\HH}{\mathcal{H}}
\numberwithin{equation}{section}

\journal{arXiv}
\begin{document}

\begin{frontmatter}



\title{A blended approach for evolving phase fields using peridynamics: Cyclic loading in quasi-brittle fracture}


\author[1]{Hayden Bromley}
\author[1]{Robert Lipton}

\affiliation[1]{organization={Department of Mathematics, Louisiana State University},
            addressline={Lockett Hall}, 
            city={Baton Rouge},
            postcode={70803-4918}, 
            state={LA},
            country={USA}}

\begin{abstract}
A field theory is presented for predicting damage and fracture in quasi brittle  materials  incorporating effects of irreversible (plastic) deformation as well as elastic moduli that soften with damage. The new observation made here is that material degradation models  consistent with plastic dissipation 
can be described by a two-point history-dependent phase field.
This approach blends a two-point phase
field with the elastic and plastic deformation evolving according to Newton's second law. Here the motion of material and the material properties (plasticity and damage) are coupled through a constitutive law non-locally in space-time.
The strain is given by an additive decomposition of elastic strain and irreversible  strain. The stress-strain behavior is described by a strength envelope and a family of unloading laws based on damage and plasticity with elastic moduli that degrade in coordination with the accumulation of irreversible strain. 
The material displacement field is  uniquely determined by the initial boundary value problem.  The theory  satisfies energy balance, with positive energy dissipation rate  in accordance with the laws of thermodynamics. 
The fracture energy of flat cracks is recovered directly from the evolution equation and is the product of energy release rate and the crack area, moreover  this formula is independent of the  length scale of non-locality. 
The formulation delivers a mesh free method for predicting crack patterns in quasi-brittle media and  simulations show quantitative and qualitative agreement with experiments, including hysteresis and damage associated with three-point bending tests on concrete and size effects for quasi-brittle materials. We conduct numerical experiments exhibiting the competition between plastic yield strength and Griffith strength on fracture nucleation. This work has been published in the  {\em Journal of the Mechanics and Physics of Solids}, 216 (2026) 106771.


\end{abstract}


\end{frontmatter}



\section{Introduction}
\label{sec.intro}

The  identification of a suitable field theory from which to pose the fracture initial boundary value problem (IBVP) is a longstanding problem of continuum mechanics.
Ideally the IBVP should deliver an  evolution for the displacement of the material inside a sample $\Omega$  based upon Newton's second law. From this a displacement damage-fracture pair is sought. The evolution is informed by initial values for the displacements and velocities and driven by boundary conditions that change in time.  The associated numerical methods for obtaining the solution follow on discretization  of the IBVP. During the past fifty years there has been conceptual and theoretical progress in understanding this challenge along several directions by physicists and engineers as well as a growing number of applied mathematicians, see 
 \cite{Freund2}, \cite{Willis},  
   \cite{XuNeedeleman1994}, \cite{marder}, \cite{OrtizPandolfi},
 \cite{silling2000reformulation}, \cite{Gao}, \cite{Nicas}, \cite{DuTaoTian}, and \cite{DalMasoLarsonToader}. 
These citations while not exhaustive do provide important developments in the theory and computation of crack patterns employing an evolution equation based exclusively on Newton's second law.   

The field theory expressed in this paper uses a peridynamic interpretation of Newton’s second law, \cite{silling2000reformulation}, \cite{silling2007peridynamic}, to evolve the displacement of the material and there is no separate equation for phase field evolution. Phase fields are introduced as damage parameters as part of a history dependent constitutive law determined by the displacement field. 
The dynamics can be articulated as an evolution on the product space $\Omega\times\Omega$  of strains between pairs of points $\yy$ and $\xx$ inside $\Omega$, see  \cite{lipton2024energy}.  A constitutive law for the relationship between force and strain between two points is provided in terms of the phase field,  see \cite{lipton2024energy}, 
\cite{coskundamerchelilipton}. The dynamic evolution of the displacement autonomously selects the pairs of points for which there is no longer force exerted between them. When the pairs are aligned transverse to a surface we recognize the surface as a new component of boundary, i.e., a crack. The energy necessary to form a crack follows directly from the evolution equation via geometric measure theory, \cite{lipton2024energy} and is precisly the fracture energy given by the critical energy release rate multiplied by the surface area of the crack.

Non-monotonic loading of concrete and geologic material require the constitutive modeling to capture both elastic unloading and reloading of the material as well as allow for strain localization and the emergence of fracture.  These phenomena have physical aspects of both elastic degradation and the accumulation of irreversible strain as revealed through experiment see for example \cite{chen2023fracture}, \cite{jenq1988mixed}. The effects of open micro-cracks  are important and are reflected in the degradation of elastic properties \cite{Budianski}, \cite{Hashin}. When compressed the specimen experiences friction from sliding motion across micro-crack faces in contact and  this behavior manifests itself as plasticity.  Recent micro-mechanics  approaches linking micro-crack motion with damage and plasticity can be found in \cite{ZhyShaoKobdo} and \cite{Uhloa}.  
The combination of these aspects were identified early on by \cite{Bazant1979} as exemplified in the work of \cite{Bezant1992}, \cite{lubliner}, \cite{Fenves}, \cite{Grassel1}, \cite{Grassell1bis}, \cite{Grassel2}, and \cite{VTK}.

Motivated by the damage mechanics set forth in \cite{fremondNedjar} we work at the mesoscale to  describe the  effects of damage and plasticity. We make make use of  two phase fields (damage fields) one associated with tension and another with compression. These phase fields are coupled to the evolution of plastic tensile and compressive strain through unloading ratios introduced by \cite{Grassel1}, \cite{Grassel2}, see Definition \ref{reversetoirreverse} and Sections \ref{Generalization} and \ref{unloadingratio}. The constitutive laws relating the force between two points are nonlocal and as mentioned depend on the time history of the strain between two points $\yy$ and $\xx$. The nonlocal forces are averaged over a ball centered at $\xx$  of small radius $\epsilon$ relative to the specimen size and the resultant force defined at $\xx$ is the macroscopic property. For symmetric domains and monotonic applied loads the solution of the nonlocal IVBP converges to the elastic wave equation coupled to a straight crack as $\epsilon\rightarrow0$, see \cite{liptonjha2021}. 
In this paper we are interested in material failure hence our blended constitutive model remembers only extreme events, thus only the maximum extension and compression of a bond up to the present time provides the memory needed at mesoscopic length scales. In this way we use the history of extreme events and avoid the {\bf overhead cost of time integration} used in plasticity and damage models. It is through meso-scopic dynamics that we numerically recover the macroscopic features of  dynamics exhibited in experiment. This aspect is mathematically described as the horizon goes to zero in \cite{lipton2016cohesive}, see also \cite{mesoscopic}. One can think of these results as dynamic homogenization tools for describing fracture models.

The approach given here extends the blended approach introduced for fracture in \cite{lipton2024energy}, and 
\cite{coskundamerchelilipton}. It applies new time dependent constitutive modeling that describes the effects of irreversible strain and damage working in concert with cyclic tension-compression loading.
The method makes exclusive use of Newton's second law and differs from traditional phase field and peridynamic methods. 
Traditional phase field methods couple a time evolved phase field with an time evolution equation for the displacement. The first such approach is formulated as  a phase field displacement system with the phase field evolving according to a  Ginzburg-Landau equation and the displacement evolution based on  Newton's second law. In this system the phase field equation is driven by the displacement field  and the displacement equation is coupled to the phase field, see \cite{Aranson,Karma}.  The second type of phase field methodology is originally developed for quasi static fracture and the displacement is coupled to the phase field through the \cite{AmTorelli} energy,  see \cite{Bourdin-Francfort-EtAl-2000}.  These phase field methods are extended to the dynamic case  and the displacement field is driven by phase field coupled through Newton's second law and phase field evolutions are driven by the minimization of an Ambrisio-Torterelli energy \cite{BourdinLarsenRichardson,LarsinOrtinerSuli}.  
Subsequently  a thermodynamically consistent  phase field method is developed using Newton's second law employing a second evolution equation  for the phase field based again on an Ambrosio-Tortorelli energy  \cite{borden2012phase}.  We point out that the methodology  presented here recovers the same dynamic fracture results obtained in   \cite{borden2012phase} for brittle fracture evolution  see  \cite{coskundamerchelilipton}.

In peridynamics the bond force between two points is referred to as the force exerted on $\yy$ by $\xx$. The constitutive law relates the bond force to the elastic strain between $\yy$ and $\xx$.
The new modeling established in this paper introduces two history-dependent phase fields one describing elastic bond force decrease (damage) in tension  and the other describing bond damage in compression.  The zones of damage localization and  fracture are  found from the phase fields as determined by history dependent constitutive modeling and Newton's second law. The material degradation described by the phase fields evolve  consistently with the plastic strain accrued by the material.  
In this approach the evolution of the two-point phase fields and plastic strain are part of the evolving constitutive law, handling plasticity and fracture evolution from quasi-static to dynamic loading. The main feature being that the {\em constitutive law evolves with the displacement providing the {\bf\em{necessary}} two way coupling between displacement and material behavior}. In this way we can describe the effects of plasticity, damage, hysteresis and  their localization onto surfaces in terms of a well posed initial value problem.

 The evolving constitutive law relates the bond force to the elastic strain between $\yy$ and $\xx$ at time $t$.  We show that the constitutive law of \cite{coskundamerchelilipton} is an edge case of the more complex time dependent constitutive law developed here that address the combined effects of irreversible strain and elastic softening (damage) for both tension and compression in sections  \ref{sec:nonlocal:model} and \ref{Generalization} of this paper.
The new  constitutive laws provide a strong form field theoretic approach from which a numerical implementation is realized as  a discretization of the continuum equations.  

All power and energy balance laws follow from the strong form of the evolution equations on multiplying the equation by the velocity and integration over the domain. 
A careful derivation recovers power and energy balance with a positive energy dissipation rate  consistent with  thermodynamics, see sections \ref{powerbalanceprocesszone2}, \ref{energybalanceprocesszone2}.  As a consequence it is found that the failure energy for  flat cracks can be calculated explicitly. For this case, the failure energy is identified as the product of two factors: one providing the explicit formula for the energy release rate, and the other giving the crack length, see section \ref{criticalenergyrelease}.  It is emphasized that in this approach these  properties are  not postulated but  instead are consequences of the constitutive law and evolution equation and follow directly from integration by parts, see section \ref{powerbalanceprocesszone2}.
A mesh free numerical method then follows for numerically predicting crack patterns using discrete approximations of the proposed continuum model.

The nonlocal method given here is motivated by the differentiable  peridynamic constitutive laws given in \cite{lipton2014dynamic,lipton2016cohesive}, and \cite{lipton2019complex}.  For these cases one can  show that any sequence of solutions associated with vanishing non-locality have subsequences converging to cluster points that are solutions of the dynamic Navier equation away from cracks (see \cite{lipton2014dynamic,lipton2016cohesive}). Additionally one can develop  methods  originally used to solve De Georgi's  conjecture  on non local approximations to the Mumford Shaw functional, see \cite{gobbino1998finite} to show the associated peridynamic energy $\Gamma$-converges to the sum of the elastic energy and the Griffith fracture energy as shown in \cite{lipton2014dynamic,lipton2016cohesive}. The elastic interaction between the  crack and surrounding peridynamic field converges to the classic zero traction condition on the crack lips, this is demonstrated for straight cracks in \cite{liptonjha2021}. The crack tip driving force is found to match the product of the energy release rate and crack tip velocity in the limit of vanishing non-locality.
This approach bypasses Mott's hypothesis, and the result emerges directly from the evolution equation, see \cite{jhalipton2020}. In turn, this provides the explicit formula relating crack velocity and the energy flowing through the crack tip, see \cite{Freund2}. In summary this nonlocal method replaces derivatives with difference quotients and the Naiver operator with a nonlocal integral operator. 
Numerical experiments show the discretization of the model presented here delivers results in accord with experiment, see Section \ref{sec.numerics}.

In this and earlier blended models the elasticity, strength and fracture toughness can be prescribed independently of each other.  
For this model, only essential material properties are used and no more.  Here the bond degradation is calibrated only by the specimen strength, critical energy release rate, the material stiffness, and the unloading ratio as defined in \cite{Grassel1}, see section \ref{sec:nonlocal:model}.
The  blended model handles both brittle fracture and quasi-brittle fracture across loading conditions, from quasi-static to dynamic fracture evolution, accommodating both monotonic and cyclic loading cases. An explicit length scale $L$ that is characteristic of the evolution emerges from the material parameters and is used to calibrate the nonlocal constitutive law. The admissible values of $L$ are constrained by the shape of the strength envelope described in the next section. The ratio of fracture toughness to material strength is naturally linked to $L$, see Section \ref{criticalenergyrelease}. 

The blended models are calibrated to the elastic moduli, fracture toughness, and strength  {\bf{differently}} than models used in  peridynamics. In the blended approach the calibration of fracture toughness and elasticity coefficients follow directly from  dimensional analysis and geometric measure theory, and the material strength is calibrated as an independent parameter as in \cite{Carpinteri}, \cite{HillerborgPetersson1976}, \cite{XuNeedeleman1994}, \cite{OrtizPandolfi}. Additionally we introduce the unloading ratio $\beta$ providing the ratio between plastic strain associated with elastic softening and irrecoverable strain as introduced by \cite{Grassel1}, \cite{Grassel2}. The value of the unloading ratio and choice of constitutive law enables one to model both brittle fracture and quasi-brittle fracture within the blended framework. Most importantly in our framework  the length scale of non-locality ``$\epsilon$,'' {\em is not a material property and is not used in the calibration of strength, fracture energy, elastic or plastic properties}. The blended model is  understood as a  well-posed approximation theory under which the effects of damage and plasticity can localize onto surfaces, see Figures \ref{P} and \ref{fig:plastic-strain-contour}.  The length scale of localization is controlled by $\epsilon$.

The peridynamic-phase field methodology developed here can be applied to quasi-brittle materials including geomaterials, masonry, and  concrete. As a specific application we consider tensile failure in concrete. We discretize the blended model and are quantitatively able to capture the hysteresis associated with crack mouth opening versus load for cyclic three-point bending experiments using our model, see Section \ref{sec.numerics}.
Notably, the deterministic size effect \cite{BazantPlanas1998}, \cite{Carpinteri},  emerges directly from our initial boundary value problem and is observed in numerical simulations, consistent with size-effects reported by \cite{hobbs} using an exponential peridynamic potential, see Section \ref{sec.numerics}. Here the numerical method used to obtain solution is meshfree in that there are no elements or geometrical
connections between the nodes and discretization is based on quadrature as in \cite{sillingaskari2005}. 

Alternatively we may study yield strength versus Griffith fracture criterion as a function of crack size.
 We consider the following numerical experiment carried out in Section \ref{strengthstability}. In this experiment we explore the reaction force of a rectangular sample undergoing the single edge notch test.  Here we load the specimen with a pre-crack in tension precisely to the onset of quasi-brittle fracture for various lengths of pre-crack. For small pre-cracks relative to the sample size the numerical experiment shows that a specimen begins to fracture when the reaction force is consistent with a homogeneous macroscopic stress close to the critical tensile yield stress of the material. For longer pre-cracks the numerical experiment shows that the crack begins to grow according to Griffith's criterion. This behavior of yield strength versus Griffith fracture criterion is consistent with  the size effect \cite{BazantPlanas1998} as well as the recent work of \cite{KumarFrancfortPalmes2}.

\bigskip

To summarize the new ideas and methods introduced in this paper are:
\begin{itemize}
\item A field theory guaranteeing existence and uniqueness of an elastic-plastic displacement in the presence of compressive-tensile cycling and  one recovers the stress field a-posteriori from the displacement field solution of the IBVP, see Sections \ref{Generalization} and \ref{wellposedness}.  
\item Calculation shows positivity of the damage and plastic dissipation in accord with the second law of thermodynamics, see Section \ref{powerbalanceprocesszone2}.
\item Plastic dissipation in tension and compression are described by a pair of two-point history-dependent phase fields, see Section \ref{Generalization}.
\item Calibration of energy release rate, elastic stiffness and strength is independent of peridynamic horizon, (length scale of non-locality), see Section \ref{sec.Straight Cracks}.
\item The novel calibration enables  the blended method to reveal competition between plastic yield strength and Griffith strength for cracks of different sizes in accord with experiment, see Section \ref{strengthstability}.
\item Discretization of evolution equation delivers a numerical method capable of reproducing hysteresis  and simultaneous crack growth, see Sections \ref{sec. hysterisiscrackgrowth} and \ref{offcenter}.
\item  Comparison of  simulations with experiments show quantitative  and qualitative agreement with three point bending tests on concrete, mixed mode cracking of L-shaped domains, and sample size effects see Sections \ref{mixedmode}, \ref{sizeeffects}, and  \ref{strengthstability}
\end{itemize}

It is noted that earlier  work \cite{She-Mo-Sh} provides peridynamic based algorithms for cyclic loading with damage and plastic effects. These algorithms are guided by constitutive models proposed for the study of concrete. Unfortunately the constitutive relations given by (15) and (16) of that paper exhibit post peak bond stiffening for a range of plastic bond strains. This is shown in the Appendix. The recent work of \cite{HuLi1}, \cite{HuLi2} provides a formal continuum mechanics based approach and considers material damage using peridynamics, see also \cite{sillinglehoucq2010}.

The paper is organized as follows: The initial boundary value problem for the fracture evolution of quasi-brittle material is formulated in the next two sections. Section \ref{wellposedness} establishes existence and uniqueness of the evolution problem using fixed-point methods for differential equations posed in  Banach space.
In section \ref{powerbalanceprocesszone2} we multiply the evolution equation by the velocity and integrate by parts to identify the damage power $\dot{\mathcal{D}}(t)$ and the elastic power $\dot{\mathcal{E}}(t)$.  This delivers a power balance law  that  is consistent with thermodynamics. In section \ref{energybalanceprocesszone2} we integrate the stress power in each bond with respect to time and then integrate in space with respect to the endpoints of the bonds $\yy$ and $\xx$. This gives us the forms of the energies $\mathcal{E}(t)$ and $\mathcal{D}(t)$ and the energy balance. In Section \ref{sec.Straight Cracks}, it is shown that the model calibration follows directly from the two-point interaction potential and energy balance and given values of elastic moduli, strength, energy release rate and unloading ratio. The IBVP delivers a mesh free numerical method for predicting crack patterns and  the results of our numerical experiments are presented in Section \ref{sec.numerics}.



\section{A nonlocal phase field formulation incorporating irreversible tensile strain and elastic compression}
\label{sec:nonlocal:model}

The body containing the damaging material $\Omega$ is a bounded domain in two or three dimensions.  A dimensional analysis is carried out and a characteristic length scale $L$ associated with the evolution is determined  in Section \ref{criticalenergyrelease}.  This length scale depends explicitly on the ratio of fracture toughness and strength. Nonlocal interactions between a point $\xx$ and its neighbors $\yy$ are confined to the sphere (disk) of radius $\epsilon$ denoted by $\HH_\epsilon(\xx)=\{\yy:\,|\yy-\xx|<\epsilon\}$. The length scale $\epsilon$ is taken  smaller than $L$ and small enough to resolve the process zone.  
 Here $V^\epsilon_d=\omega_d\epsilon^d$ is the $d$-dimensional volume of the ball $\HH_\epsilon(\xx)$ centered at $\xx$ where $\omega_d$ is the volume of the unit ball in $d$ dimensions.  The displacement of a point $\xx$ inside the material at time $t$ is $\uu(t,\xx)$ and is defined for $0\leq t\leq T$ and $\xx$ in $\Omega$. We write $\uu(t)=\uu(t,\cdot)$ and introduce the 
two-point strain $S(\yy, \xx, \uu(t))$ between the point $\xx$ and any point $\yy \in \HH_\epsilon(\xx)$  resolved along the direction $\ee$ given by
\begin{align}\label{strain}
    S(\yy, \xx, \uu(t)) = \frac{(\uu(t,\yy)-\uu(t,\xx))}{|\yy-\xx|}\cdot\ee, & \hbox{  where  } \ee=\frac{\yy-\xx}{|\yy-\xx|},
\end{align}	
and introduce the scaled strain
\begin{align}\label{def of r}
r:=r(\yy,\xx,\uu(t))=\sqrt{\frac{|\yy-\xx|}{L}}S(\yy,\xx,\uu(t)).
\end{align}
The strain satisfies the symmetry $S(\yy, \xx, \uu(t)) = S(\xx, \yy, \uu(t)) $.

Displacement loads are prescribed on the boundary of the specimen $\Omega$. For the nonlocal model, the displacement $\UU(t,\xx)$ is prescribed on a layer $\Omega_D^\epsilon$ surrounding part of the specimen see, e.g., \cite{DuTaoTian}. We write $\Omega^*=\Omega\cup\Omega_D^\epsilon$ where the layer is of maximum thickness greater than or equal to $\epsilon$, see Figure \ref{Dirichlet} . With this in mind, we write
\begin{equation}\label{scale2}
\rho^{\epsilon}(\yy,\xx)=\frac{{\chi_{\Omega^*}(\yy)}J^\epsilon(|\yy-\xx|)}{ \epsilon V^\epsilon_d},
\end{equation}
where {$\chi_{\Omega^\ast}$} is the characteristic function of $\Omega^\ast$, $J^\epsilon(|\yy-\xx|)$ is the influence function, a positive function on the ball of radius $\epsilon$ centered at $\xx$ and is radially decreasing taking the value $M$ at the center of the ball  and $0$  for $|\yy-\xx|\geq\epsilon$. The radially symmetric influence function is written $J^\epsilon(|\yy-\xx|)=J(|\yy-\xx|/\epsilon)$. The nonlocal kernel is scaled by $\epsilon^{-1}(V_d^{\epsilon})^{-1}$ enabling the model to be characterized by, 1) the  linear elastic shear $G$ and Lam\'e $\lambda$ moduli  in regions away from the damage,  2) the critical energy release rate ${G}_c$ and, 3) the  strength of the material, this is done in Section \ref{sec.Straight Cracks}.

\subsection{Constitutive laws with memory: Initial Boundary Value problem for Quasi-Brittle fracture incorporating irreversible strain and damage under loading. }
\label{sec.Constiutivelaws BVP}

\begin{figure}
    \centering
    \begin{subfigure}{.45\linewidth}
       \begin{tikzpicture}[scale=0.7]

        \draw[<->,thin] (-1,0) -- (8,0);   
        \draw[<->,thin] (0,-2) -- (0,5);    
        \node[left] at (0.9,4.5) {$\sigma$};   
        \draw[thick] (2,0)--(2,-0.2);        
        \draw[thick] (7,0)--(7,-0.2);        
        \draw[thick] (0.98,0)--(0.98,-0.2);        
        \node[below] at (2,-0.2) {$S_t^C$};   
        \node[below] at (7,-0.2) {$S_t^F$};   
            \node[below] at (1.1,-0.2) {$P^\ast$};   
        \node[right] at (8,0) {$S$};        
        \draw [dashed,thick] (2,3.5) to [out=-35,in=180] (7,0.0); 
L


        \node[right] at (3,3) {$(g^+)'(r^*)/\sqrt{|\yy-\xx|/L}$};   
        \draw[dark green,ultra thick] (0,0)--(2,3.5);
        \draw[<->,dark green,ultra thick] (0.4,0.7)--(1.6,2.8);
        \draw[blue ,ultra thick] (-1,-1.75)--(0,0);
        \draw[dark green,ultra thick] (1,1.75)--(1.2,1.75)--(1.2,2.1); 
        
        \node[right] at (1.2,1.75) {\color{dark green}{$\bar{\mu}$}};

        \draw[blue ,ultra thick] (0,-1.75)--(1,0);
        \draw[blue,ultra thick] (1,-1.75)--(2,0);
        \draw[blue,ultra thick] (2,-1.75)--(3,0);
        \draw[blue,ultra thick] (3,-1.75)--(4,0);
        \draw[blue,ultra thick] (4,-1.75)--(5,0);
        \draw[blue,ultra thick] (4.5,-1.75)--(5.5,0);
        
        \draw[dark green,ultra thick] (1,0)--(2.95,2.7);
        \draw[<->,dark green,ultra thick] (1.39,0.54)--(2.56,2.16);
        
        \draw[dark green,ultra thick] (2,0)--(3.7,1.9);
        \draw[dark green,ultra thick] (3,0)--(4.4,1.2);
        \draw[dark green,ultra thick] (4,0)--(5,0.65);
        \draw[dark green,ultra thick] (5,0)--(5.8,0.2);
        \draw[dark green,ultra thick] (1.975,1.35)--(2.175,1.35)--(2.175,1.6269); 
        \node[right] at (2.175,1.35){\color{dark green}{$\gamma\bar{\mu}$}};
        
        \draw[->,red,ultra thick] (5.5,0)--(8,0);
        
        \draw[thick] (5.5,0)--(5.5,-0.2);
        \node[below] at (5.65,-0.3) {\color{red}{$\beta S_t^F$}};
        
        \end{tikzpicture}
		  \caption{}
		  \label{ConvexConcavea}
    \end{subfigure}
    \hskip2em
    \begin{subfigure}{.45\linewidth}
        \begin{tikzpicture}[scale=0.7]

\draw[thin] (-1,0) -- (8,0);   
\draw[thin] (0,-2) -- (0,5);    
\node[right] at (0,4.5) {$\sigma$};   
\node[right] at (3.0,4.1) {$(g^-)'(r_*)/\sqrt{|\yy-\xx|/L}$};   
\draw[thick] (2,0)--(2,-0.2);        
\draw[thick] (2.6,0)--(2.6,0.2);        
\draw[thick] (7,0)--(7,-0.2);        

\node[below] at (2.1,-0.2) {$S_c^Y$};   
\node[above right] at (2.5,-0.0) {$S_c^C$};   
\node[below] at (7.2,-0.2) {$S_c^F$};   

\draw [dashed,thick] (2,3.5) to [out=55,in=180] (7,0.0); 

\draw[dark green,ultra thick] (-1,-1.75)--(0,0);
\draw[blue,ultra thick] (0,0)--(2,3.5);
\draw[<->,blue,ultra thick] (0.4,0.7)--(1.6,2.8);

\draw[blue,ultra thick] (1,1.75)--(1.2,1.75)--(1.2,2.1); 

\node[right] at (1.2,1.75) {\color{blue}{$\bar{\mu}$}};

\draw[dark green,ultra thick] (0,-1.75)--(1,0);
\draw[dark green,ultra thick] (1,-1.75)--(2,0);
\draw[dark green,ultra thick] (2,-1.75)--(3,0);
\draw[dark green,ultra thick] (3,-1.75)--(4,0);
\draw[dark green,ultra thick] (4,-1.75)--(5,0);
\draw[dark green,ultra thick] (5,-1.75)--(6,0);

\draw[blue,ultra thick] (1,0)--(3.5,3.3);

\draw[blue,ultra thick] (2,0)--(4.2,2.4);
\draw[blue,ultra thick] (3,0)--(4.78,1.58);
\draw[blue,ultra thick] (4,0)--(5.3,0.95);
\draw[blue,ultra thick] (5,0)--(5.95,0.35);
\draw[blue,ultra thick] (2.04,1.35)--(2.25,1.35)--(2.25,1.65); 
\node[right] at (2.175,1.48){\color{blue}{$\gamma_-\bar{\mu}$}};

\draw[->,red,ultra thick] (6,0)--(8,0);
\node[right] at (8,-0.0) {{$ S$}};
\draw[thick] (6,0)--(6,-0.2);
\node[below] at (6.25,-0.3) {\color{red}{$\beta_- S_c^F$}};

\end{tikzpicture}
        \caption{}
		\label{ConvexConcaveb}
    \end{subfigure}
    \caption{  (a) 
     Tensile stress vs strain curve sketched for bond strain with plastic yielding only after peak stress is reached, i.e., $S_t^Y=S_t^C$.  Blue lines are admissible stress strain curves in compression and green lines are admissible stress strain curves in tension.  These broken lines foliate the strength domain and are admissible constitutive laws for admissible bond stress-strain pairs $(\sigma,S)$. Stress-strain pairs on the strength envelope of unloading laws $(g^+)'(r^\ast)/\sqrt{|\yy-\xx|/L}$ given by dashed curve. The strain is reversible and elastic with constant bond stiffness  until $S^*=S_t^Y=S_t^C$. For $S_t^C\leq S^*\leq S_t^F$  then bonds exhibit plasticity and elastic softening. The bond stiffness is proportional to  the phase field and corresponding unloading laws are the broken green-blue lines. The  plastic tensile strain  is $P^*$ given by \eqref{damageplastique} and is the $S$ axis  intercept of the unloading law. The stiffness decreases from $\overline{\mu}=\partial_{r}g^+(r_t^C)/r_t^C$  smoothly to $0$.  The bond offers zero tensile stiffness for bonds broken in tension (red), see Definition \ref{def: Second}. However the bond continues to elastically resist negative strains, see blue unloading line connected to red line. (b) Stress versus strain curve sketched with plastic yielding before peak stress now given for compression loading. In this figure the sign of compressive strain and stress are reversed. }
 
    \label{ConvexConcave}
\end{figure}
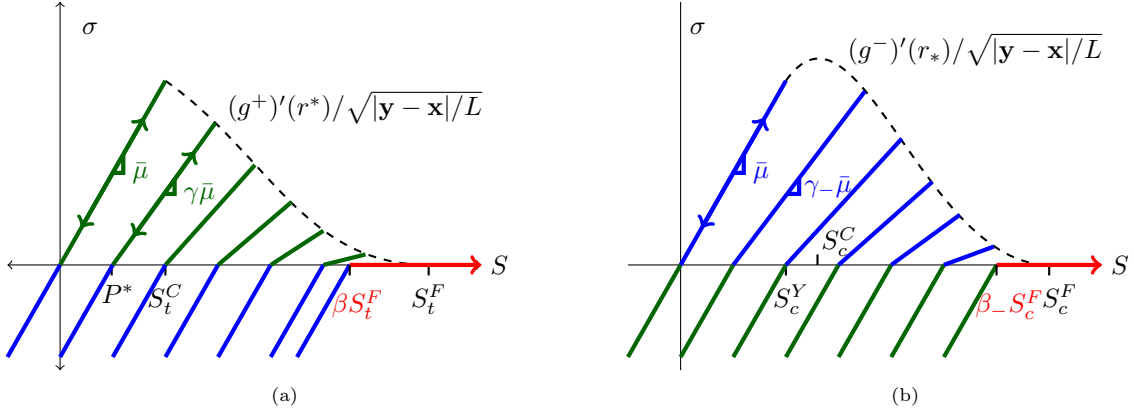
 To illustrate the ideas we begin by describing a constitutive model describing material that damages under tensile loading and to first approximation behaves elastically under compressive loading.
 The tensile strain in undamaged material is elastic while damaged material responds to force with elastic-plastic behavior given by an additive decomposition of elastic strain and irreversible  strain. For a bond in tension  the loading law for the stress is initially linear for tensile strain up to $S_t^Y$,  then yields and  goes nonlinear exhibiting strain hardening up to peak tensile stress. Here the strain $S_t^C$ corresponds to the strain at peak tensile stress, and $S_t^F$ corresponds to the tensile strain at failure. The stress-strain behavior is described by a strength envelope and a family of elastic-damage plastic unloading laws.
The force between two points is a function of the  strain. 
\begin{definition}[Bond]
\label{First}
The force associated with the interval between two points $\yy$ and $\xx$ is referred to as a bond.
\end{definition}

The bond stiffness at the strains $S_t^Y$,  $S_t^C$, and $S_t^F$ depend on the scaled bond length $\sqrt{|\yy-\xx|/L}$ and are given by  $r_t^Y$, $r_t^C$, and $r_t^F$ determined by the material
\begin{align}\label{strains}
S_t^Y=&\frac{r_t^Y}{\sqrt{|\yy-\xx|/L}}\nonumber\\
S_t^C=&\frac{r_t^C}{\sqrt{|\yy-\xx|/L}}\nonumber\\
S_t^F=&\frac{r_t^F}{\sqrt{|\yy-\xx|/L}}.
\end{align}
The square root dependence on relative bond length is introduced in
the regularized bond model \cite{lipton2014dynamic}, \cite{lipton2016cohesive} and provides the envelope for fast fracture in \cite{lipton2024energy}. It also provides the envelope for the blended model \cite{coskundamerchelilipton} for brittle and quasi-brittle materials. The rational behind the scaling is that its  required for delivering energy balance  between  the elastic,  fracture, and kinetic energies associated with the evolution equation, see Section \ref{energybalanceprocesszone2}.

We introduce the memory function given by the maximum strain up to time $t$ given by
\begin{align}
\label{scaledmax}
S^*(t,\yy,\xx,\uu)=\max_{0\leq \tau\leq t}\{ S(\yy, \xx, \uu(\tau))\}.
\end{align}
and set $$r^\ast:=r^\ast(t,\yy,\xx,\uu)=\sqrt{\frac{|\yy-\xx|}{L}}S^*(\yy,\xx,\uu(t)).$$

To shorten the exposition and when appropriate we will denote $S^*(t,\yy,\xx,\uu)$ by $S^*(t)$ and $r^*(t,\yy,\xx,\uu)$ by $r^*(t)$.
We introduce the strength envelope associated with the nonlinear potential function $g^+(r)$ given by $\partial_r g^+(r)$, see Figure \ref{ConvexConcavea}.    
The piece wise linear unloading laws foliating the strength domain are characterized by strains that are additively decomposed into elastic and plastic parts. The plastic strain is given by the $r^*$ axis intercept of the unloading law  in Figure \ref{ConvexConcavea}. Here the elastic force is linearly related to the reversible  strain, i.e., the  elastic part of the strain. Importantly damage and plasticity are coordinated in that the elasticity of the material is decreasing with increasing irreversible tensile strain. Here decreasing elasticity is associated with material damage.
On the other hand, for compressive strain the force between two points is a simple elastic force and characterized by a constant elastic stiffness.  
The strength envelope (see Figure \ref{ConvexConcavea}) is defined by
\begin{equation}\label{failure envelope}
\frac{\partial_{r^*} g^+(r^*(t))}{\sqrt{{|\yy-\xx|}/L}}.
\end{equation}
On the strength envelope, the present strain at time $t$ is the maximum strain and  $r^*=r$. 
As an example, consider the simple bilinear model where $r_t^Y=r_t^C$ defined for bonds in tension given by
\begin{equation}\label{bilinear}
\partial_{r^*} g^+(r^*)=\left\{\begin{array}{lc}\overline{\mu}r^*,& r^*< r_t^C,\\
\overline{\mu}r_t^C\frac{r_t^F-r^*}{r_t^F-r_t^C},&r_t^C\leq r^*\leq r_t^F,\\
0,&r_t^F<r^*.\end{array}\right.
\end{equation}

We introduce the phase field $\gamma(t,\yy,\xx,\uu)$ that describes  softening  of elastic moduli with  $0\leq\gamma(t,\yy,\xx,\uu)\leq 1$. When the bond is elastic  with no irreversible strain and no elastic softening then $\gamma(t,\yy,\xx,\uu)=1$.  At the other extreme when the bond is fully damaged with  maximum  irreversible strain, consistent with complete elastic softening, then $\gamma(t,\yy,\xx,\uu)=0$. To begin we suppose no elastic softening occurs so bond stiffness remains at $\bar{\mu}$ and only plastic strain is accrued by the bond under tensile loading and this strain is denoted by $P^*_i(t,\yy,\xx,\uu)$. For this case the maximum  remaining recoverable strain associated with a point  on the strength envelope $(S^*,\partial_{r^*}g^+(r^*)/\sqrt{|\yy-\xx|/L})$ is 
\begin{align}\label{maxrecoverablestrain}
  \frac{\partial_{r^*}g^+(r^*(t,\yy,\xx,\uu))}{\overline{\mu}\sqrt{{|\yy-\xx|}/L}},
\end{align}
so  the plastic tensile  strain  without elastic  softening is  given by
\begin{align}
    \label{plastique}
    P^*_i(t,\yy,\xx,\uu)=S^*(t,\yy,\xx,\uu) -\frac{\partial_{r^*}g^+(r^*(t,\yy,\xx,\uu))}{\overline{\mu}\sqrt{{|\yy-\xx|}/L}}\geq 0.
\end{align}
Similarly the plastic strain that takes into account material softening is expressed in terms of the phase field $\gamma(t,\yy,\xx,\uu)$ and given by
\begin{align}
    \label{damageplastique}
    P^*(t,\yy,\xx,\uu)=S^*(t,\yy,\xx,\uu) -\frac{\partial_{r^*}g^+(r^*(t,\yy,\xx,\uu))}{\overline{\mu}\gamma(t,\yy,\xx,\uu)\sqrt{{|\yy-\xx|}/L}}\geq 0.
\end{align}

To complete the definition of the phase field we introduce the unloading parameter characterizing the combination of degradation in elastic moduli and plastic strain associated with the material \cite{Grassel1}. 
\begin{definition} {\bf The unloading ratio $\beta$.}\label{reversetoirreverse}\\
The parameter $\beta$ is the ratio of plastic strain that takes into account material softening   to the plastic strain strain  in the absence of material softening and $0\leq \beta\leq 1$. 
\end{definition}

The explicit formula for the phase field follows from the definition of $\beta$ and is given by 
\begin{align}
    \label{ratio}
    \beta=\frac{P^*(t,\yy,\xx,\uu)}{P^*_i(t,\yy,\xx,\uu)}. 
\end{align}
Here $\beta$ is a coarse scale measure of microscale physics taking place on length scales below those resolved by the IBVP. The multiscale nature of the $\beta$ parameter is discussed in section \ref{unloadingratio}.
  This choice of $P^\ast(t,\yy,\xx,\uu)$ based on $\beta$ insures that $0\leq \gamma(t,\yy,\xx,\uu)\leq 1$ and Lipchitz continuous in a suitable Banach space norm as shown in Section \ref{wellposedness}. Rearranging terms in \eqref{ratio}  gives the explicit formula for the phase field
\begin{align}
    \label{gamma}
    \gamma(t,\yy,\xx,\uu))=\overline{\mu}^{-1}\frac{\partial_{r^*}g^+(r^*(t,\yy,\xx,\uu))}{(S^*-\beta P^*_i(t,\yy,\xx,\uu))\sqrt{|\yy-\xx|/L}},  & \hbox{ for $\beta$ fixed and chosen in $0\leq \beta\leq 1$.}
\end{align}
For any fixed value of $0\leq \beta<1$  the phase field $\gamma$ is one for an undamaged bond corresponding to $-\infty<S^\ast(t,\yy,\xx,\uu)<S_t^Y$, becomes decreasing with respect to increasing irreversible strain for $S_t^Y\leq S^\ast(t,\yy,\xx,\uu) \leq S_t^F$  and zero for a fully damaged bond corresponding to $S_t^F\leq S^\ast(t,\yy,\xx,\uu)$.  For $-\infty<S^\ast(t,\yy,\xx,\uu)<S_t^Y$ one has $P^{\ast}(t,\yy,\xx,\uu)=0$ and this corresponds to  the numerator and denominator  of \eqref{gamma}  taking the common value $\overline{\mu}S^\ast(t,\yy,\xx,\uu)\sqrt{|\yy-\xx|/L}$ and $\gamma(t,\yy,\xx,\uu)=1$. The graph of $\gamma(t,\yy,\xx,\uu)=\gamma(\sqrt{|\yy-\xx|/L}S^\ast(t,\yy,\xx,\uu))$ is shown in {Figure \ref{ConvexConcaveC}. 
The edge case $\beta=0$ represents softening of elastic moduli under loading with no plastic strain being accrued and all unloading curves return to the origin, this case was considered  in \cite{coskundamerchelilipton}.  The edge case given by $\beta=1$, corresponds no softening of elastic moduli under loading and increasing plastic strain.
Values $0<\beta\leq1$ deliver softening elastic moduli under loading and increasing plastic strain. 
The condition $0\leq \beta\leq 1$
insures all stress strain pairs inside the evolving strength domain belong to a unique unloading law, see Figure  \ref{ConvexConcavea}.

To expedite the presentation we set $\gamma(t):=\gamma(t,\yy,\xx,\uu)$ when representing the phase field associated with a bond between $\yy$ and $\xx$.
Note $\gamma(t)$ is constant on unloading curves and  $\dot{\gamma}(t)< 0$ only when the unloading curve intersects the strength envelope and $r^*(t,\yy,\xx,\uu)>r_t^Y$. The time corresponding to $r^*(t,\yy,\xx,\uu)=r_t^Y$ is denoted by $t^Y_{\yy,\xx}$ and the time $r^*(t,\yy,\xx,\uu)=r_t^F$ is denoted by $t^F_{\yy,\xx}$,  and we abbreviate the notation and write $t^Y$ and $t^F$ when there is no chance of confusion.  A  prototypical phase field profile is given in Figure \ref{ConvexConcaveC}.

It is evident that $\dot{P}^*(t,\yy,\xx,\uu)$ is non-negative, since
\begin{align}\label{plastic44}
\dot  {P}^*(t,\yy,\xx,\uu)=\beta\left(\dot{S}^*(t,\yy,\xx,\uu)-\frac{\partial_t\partial_{r^*} g^+(r^*(t,\yy,\xx,\uu))}{\overline{\mu}\sqrt{{|\yy-\xx|}/L}}\right)\geq 0. 
\end{align}
The elastic strain  ${E}(\yy,\xx,\uu(t))$ is  given by
\begin{align}
    \label{elastic}
    {E}(\yy,\xx,\uu(t)):=S(\yy,\xx,\uu(t))-P^*(t,\yy,\xx,\uu).
\end{align}
Using \eqref{damageplastique} one easily checks that 
\begin{align}
    \label{check}
    \overline{\mu}\gamma(t)\left(S^*(t,\yy,\xx,\uu)-P^*(t,\yy,\xx,\uu)\right)= \frac{\partial_{r^*} g^+(r^*(t,\yy,\xx,\uu))}{\sqrt{{|\yy-\xx|}/L}},
\end{align}
and we  point out that when $P^\ast(\yy,\xx,t,\uu)=0$ the bond is in the linear elastic regime.

If $(\overline{\mu}\gamma(t){E}^*(\yy,\xx,\uu(t))$ is the current stress on the intersection of the unloading curve and strength envelope and  $\overline{\mu}\gamma(t){E}(\yy,\xx,\uu(\tau))$ is any other stress belonging to  the same unloading curve at some time $\tau$ then from \eqref{plastic44} we get
\begin{align}
\label{maxdiss}
\left(\overline{\mu}\gamma(t){E}^*(\yy,\xx,\uu(t))-\overline{\mu}\gamma(t){E}(\yy,\xx,\uu(\tau))\right)\dot{P}^\ast(t,\yy,\xx,\uu)\geq 0,
\end{align}
which is precisely the postulate of maximum plastic dissipation proposed independently by \cite{Mises},  \cite{Taylor}, and \cite{Hill}, but now in the context of damage and plasticity.

We assume the material resists compressive strain elastically and we write the elastic stiffness defined for tensile and compressive strain  as
\begin{equation}\label{extendmugamma}
\overline{\mu}\gamma(t):=\left\{\begin{array}{lc} \overline{\mu}\gamma(t),&E(\yy,\xx,\uu(t))> 0,\\
\overline{\mu}, & E(\yy,\xx,\uu(t))< 0.
\end{array}\right.
\end{equation}
The constitutive law relating bond stress $\sigma$ to bond strain $S(\yy,\xx,\uu(t))$ is continuous in $t$ for $\yy$, $\xx$ fixed\footnote{more precisely a.e. for $(\yy,\xx)$ in $\Omega\times\Omega$} and described by
\begin{equation}\label{sigmatensile}
{\sigma}(t,\yy,\xx,\uu):=\left\{\begin{array}{lc} \overline{\mu}\gamma(t)E(\yy,\xx,\uu(t)),& E(\yy,\xx,\uu(t))> 0,\\
0, &  E(\yy,\xx,\uu(t))= 0\\
\overline{\mu} E(\yy,\xx,\uu(t)), & E(\yy,\xx,\uu(t))< 0.
\end{array}\right.
\end{equation}

The constitutive law relating bond force per unit volume to elastic strain is described by
\begin{equation}
	\boldsymbol{f}^\epsilon(t,\yy,\xx,\uu)=\rho^\epsilon(\yy,\xx) \sigma(t,\yy,\xx,\uu)\boldsymbol{e},
	\label{Eqn.const}
\end{equation}
where $\sigma$ is defined by \eqref{sigmatensile}. The unloading curves for each pair of points $\yy,\xx$ described by \eqref{Eqn.const} are given by the broken lines as portrayed in Figure \ref{ConvexConcavea}.

 \begin{definition}[Broken Bond]
 \label{def: Second}
In this treatment, we say a bond between two points $\yy$ and $\xx$ ``fails in tension'' or ``broken in tension''  if and only if the force  between $\yy$ and $\xx$  is zero when the total strain $S(\yy,\xx,\uu(t))$ is $S^F$, i.e., from \eqref{elastic}),  when  irreversible strain is equal to $\beta S^F$ and  elastic strain is equal to $(1-\beta) S^F$.  The broken bond corresponds to  a vanishing phase field, $\gamma(t)=0$.
\end{definition}
\noindent This notion of bond failure still allows the ``bond'' to resist compressive force due to elastic strain, see Figure \ref{ConvexConcavea}.

\begin{figure}
    \centering
 
        \begin{tikzpicture}[xscale=0.8,yscale=0.8]
		    \draw [<-,thick] (0,3) -- (0,-3);
			\draw [->,thick] (-5,0) -- (3.5,0);
            \draw [-,thick] (-5,1.5) -- (1.5,1.5) to [out=-45,in=180] (3,0.0);

            \draw [-,thick] (-5,1.5) -- (1.5,1.5); 
			
			\draw (0.0,3.0) -- (0.0, 2.8);
			\node [below] at (-0.2,1.5) {${1}$};
			\draw (0.0,0.2) -- (0.0, -0.2);
			\node [below] at (-0.2,-0.2) {${0}$};
			\draw (2.8,-0.2) -- (2.8, 0.0);
			\draw (1.5,-0.2) -- (1.5, 0.0);
			\node [below] at (1.5,-0.2) {${S_t}^Y$};
			\node [below] at (2.8,-0.2) {${S_t}^F$};
			\node [right] at (3.5,0) {$S^\ast$};
            \node [right] at (0,2.2) {$\gamma(t,\yy,\xx,\uu))=\gamma(\sqrt{|\yy-\xx|/L}S^\ast(t,\yy,\xx,\uu))$};
		  \end{tikzpicture}
    \caption{ $\gamma(t,\yy,\xx,\uu)$ is one for $S^\ast\leq S_t^Y$ then decays to zero for $S_t^Y<S^\ast< S_t^F$. It is evident from the graph that $\gamma$ is Lipchitz continuous in $S^\ast$.}
    \label{ConvexConcaveC}
\end{figure}
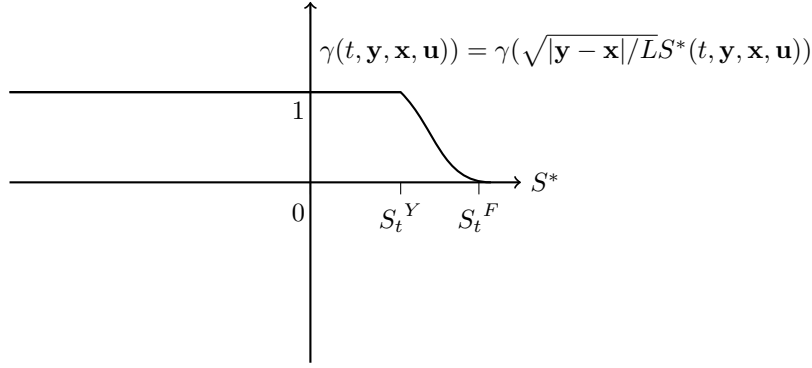


The nonlocal force density $\LL^\epsilon[\uu](t,\xx)$ defined for all points $\xx$ in $\Omega$ is  given by
\begin{align}
      \label{eq: force2}
 \LL^\epsilon [\uu](t,\xx) = -\int_{\Omega^*} \boldsymbol{f}^\epsilon(t,\yy,\xx,\uu) \,d\yy.
\end{align}
The initial boundary value problem for fracture evolution is to
find a displacement $\uu(t,\xx)$ on $\Omega^*$ that satisfies 
\begin{align}\label{eq: prescrib boundary displacement}
\uu(t,\xx)=\UU(t,\xx)& \hbox{ for }\xx \hbox{ in } \Omega^\epsilon_D,
\end{align}
and for $\xx$ in
$\Omega$  satisfies the evolution equation 
\begin{align}\label{eq: linearmomentumbal2}
\rho\ddot{\uu}(t,\xx)+ \LL^\epsilon [\uu](t,\xx) =\bb(t,\xx),
\end{align}
for $0<t<T$, and satisfies the initial conditions on the displacement and velocity given by
\begin{align}
\uu(0,\xx)=\uu_0(\xx),\nonumber\\
\dot{\uu}(0,\xx)=\vv_0(\xx).\label{initialconditions2}
\end{align}
To illustrate ideas, we fix all bonds crossing the interface between $\Omega$ and $\Omega^\epsilon_D$ to be purely elastic throughout the evolution and have stiffness $\overline{\mu}$ as well as all bonds between points in $\Omega^\epsilon_D$.

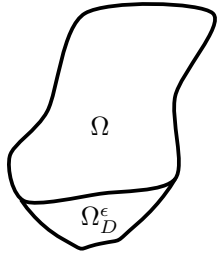
\begin{figure}
    \centering
    
      \begin{tikzpicture}[>=latex]
 \draw[line width=.5mm] plot [smooth cycle] coordinates {(0,0) (1,0.1) (2,0.3) (2,1.4) (2.5,2.5) (0.8,2.5) (0.3,1.2) (-0.2,0.6) } node at (1,1) {$\Omega$};
 \draw[line width=.5mm] plot [smooth,tension=1.2] coordinates {(-0.05,0.02) (0.5,-0.5) (1.0,-0.565) (1.5,-0.295) (2,0.3)} node at (1.0,-0.25) {$\Omega_D^\epsilon$};
\end{tikzpicture}
    \caption{ Domain $\Omega$ with prescribed Dirichlet data on $\Omega_D^\epsilon$. The union is denoted by $\Omega^*$}
    \label{Dirichlet}
\end{figure}

\subsection{Elastic Resistance to Compression Across a Crack }
\label{sec.Straight Cracks Compression}

In this model, cracks are created by bonds that offer zero stiffness under tension. On the other hand, bonds under compressive stress continue to resist elastically. When a bond ceases to resist tension the constitutive relation is illustrated by the red branch of the force-strain relation in Figure \ref{ConvexConcavea}. Let $\nn$ denote the unit normal vector perpendicular to the crack. Let $\yy-\xx$ denote all line segments that cross a planar crack such that $(\yy-\xx)\cdot\nn>0$ and $\epsilon>|\yy-\xx|$. For deformation fields such that $(\uu(t,\yy)-\uu(t,\xx))\Vert\nn$ and $(\uu(t,\yy)-\uu(t,\xx))\cdot\nn<0$, basic geometric considerations show that  the total strain is compressive, ie., $S(\yy,\xx,\uu(t))<0$ for  all bonds $|\yy-\xx|<\epsilon$.  
Since $P^*(t,\yy,\xx,\uu(t))\geq 0$ we get $E(\yy,\xx,\uu(t))<0$ for $S(\yy,\xx,\uu(t))<0$ and the constitutive relation \eqref{extendmugamma} shows that the bond force elastically opposes the compressive strain across the crack.


\section{A nonlocal phase field formulation incorporating irreversible tensile strain and compression.}
\label{Generalization}

We extend the two point constitutive law between $\yy$ and $\xx$ to include elastic-plastic damage for both  tensile strain and compressive strain. 
The memory function  used for tensile strain and the memory function used for compressible strain are given by  $S^\ast(t,\yy,\xx,\uu)=max_{\tau\leq t}\{S(\yy,\xx,\uu(\tau))\}$ and $S_{\ast}(t,\yy,\xx,\uu)=min_{\tau\leq t}\{S(\yy,\xx,\uu(\tau))\}$ respectively.
In this  context we denote the phase field in tension defined by \eqref{gamma} as $\gamma_+(\yy,\xx,t,\uu)$ and the unloading ratio in tension defined by \eqref{ratio} as $\beta^+$. Here the derivative of the potential for a bond in tension is now denoted by $\partial_{r^\ast}g^+(r^\ast)$.

For a bond in compression  the loading law for the stress is initially linear for compressive strain up to $S^Y_c$,  then yields and  goes nonlinear exhibiting strain hardening up to peak compressive stress. Here the strain $S_c^C$ corresponds to the strain at peak compressive stress, and $S_c^F$ corresponds to the compressive strain at failure.
The bond potential for compression is prescribed and given by $\partial_{r_{\ast}} g^-(r_\ast)$, see Figure \ref{ConvexConcaveb}.
The compressive potential is characterized by the compressive strains $S_c^Y$, $S_c^C$, and $S_c^F$ depends on the scaled bond length $\sqrt{|\yy-\xx|/L}$ and are given by $r^Y_c$  $r_c^C$, and $r_c^F$ determined by the material with,
\begin{align}\label{strains}
S_c^Y=&\frac{r_c^Y}{\sqrt{|\yy-\xx|/L}}\nonumber\\
S_c^C=&\frac{r_c^C}{\sqrt{|\yy-\xx|/L}}\nonumber\\
S_c^F=&\frac{r_c^F}{\sqrt{|\yy-\xx|/L}},
\end{align}
and set $$r_\ast:=r_\ast(t,\yy,\xx,\uu)=\sqrt{\frac{|\yy-\xx|}{L}}S_\ast(\yy,\xx,\uu(t)).$$
It is noted that the length scale under compression $L$ can be different than under tension.

The plastic compressive strain in the absence of elastic softening is given by
\begin{align}
    \label{plastiqque2}
    {P_{\ast,i}}(t,\yy,\xx,\uu):=\left(S_\ast(t,\yy,\xx,\uu)-\frac{\partial_{r_\ast} g^-(r_\ast(t,\yy,\xx,\uu))}{\overline{\mu}\sqrt{{|\yy-\xx|}/L}}\right)\leq 0,
\end{align}
and the plastic strain consistent with elastic softening is defined through the unloading ratio in compression by
\begin{align}
    \label{ratio2}
    \beta^-=\frac{P_\ast(t,\yy,\xx,\uu)}{P_{\ast i}(t,\yy,\xx,\uu)},  & \hbox{ $0\leq \beta^-\leq 1$.}
\end{align}
This choice of plastic strain $P_\ast(t,\yy,\xx,\uu)$ based on $\beta_-$ insures that $0\leq \gamma_-(t,\yy,\xx,\uu)\leq 1$ and is Lipchitz continuous in a suitable Banach space norm as shown in Section \ref{wellposedness}. It follows from \eqref{ratio2} that
\begin{align}
    \label{plasticc}
    {P_*}(t,\yy,\xx,\uu)&=\beta_-\left(S_*(t,\yy,\xx,\uu)-\frac{\partial_{r_\ast} g^- (r_\ast(t,\yy,\xx,\uu))}{\overline{\mu}\sqrt{|\yy-\xx|L}}\right)\nonumber\\
    &=S_\ast-\frac{\partial_{r_*} g^-(r_*(t,\yy,\xx,\uu))}{\overline{\mu}\gamma_-(t,\yy,\xx,\uu)\sqrt{{|\yy-\xx|}/L}}\leq 0.
\end{align}
From \eqref{plasticc} the compressive phase field $\gamma_-$ is defined by
\begin{align}
    \label{gammaminus}
    \gamma_-(t,\yy,\xx,\uu)=\overline{\mu}^{-1}\frac{\partial_{r_\ast}g^-(r_\ast(t,\yy,\xx,\uu))}{(S_\ast-P_{\ast}(t,\yy,\xx,\uu))\sqrt{{|\yy-\xx|}/L}}. 
    \end{align}
For any fixed value of $0\leq \beta^-<1$  the phase field $\gamma_-$ is one for an undamaged bond corresponding to $\infty>S_\ast(t,\yy,\xx,\uu)>S_c^Y$ and becomes decreasing with respect to decreasing irreversible strain in compression for $S_c^Y\geq S_\ast(t,\yy,\xx,\uu) \geq S_c^F$  and zero in compression for a fully damaged bond corresponding to $S_c^F\geq S_\ast(t,\yy,\xx,\uu)$. For $\infty>S_\ast(t,\yy,\xx,\uu)>S_c^Y$ there is zero plastic strain, i.e., $P_{\ast}(t,\yy,\xx,\uu))=0$ and this corresponds with  the numerator and denominator  of \eqref{gammaminus}  taking the common value $\overline{\mu}S_\ast(t,\yy,\xx,\uu)\sqrt{|\yy-\xx|/L}$, hence $\gamma_-(t)=1$.  Last it follows from \eqref{gammaminus} that 
\begin{align}
    \label{check2}
    \overline{\mu}\gamma_-(t)\left(S_*(t,\yy,\xx,\uu)-P_*(t,\yy,\xx,\uu)\right)= \frac{\partial_{r} g^-(r_*(t,\yy,\xx,\uu))}{\sqrt{{|\yy-\xx|}/L}},
\end{align}
    
For tensile loading outlined in the previous section, the derivitave of the bond stress potential is labeled $\partial_rg^+(r^\ast)$ and  the unloading ratio in tension is now rewritten as
\begin{align}
    \label{ratio3}
    \beta^+=\frac{P^\ast(t,\yy,\xx,\uu)}{P^{\ast}_i(t,\yy,\xx,\uu)},  & \hbox{ $0\leq \beta^+\leq 1$.}
\end{align}
As mentioned earlier this choice of $P^\ast(t,\yy,\xx,\uu)$ based on $\beta^+$ insures that $0\leq \gamma_+(t,\yy,\xx,\uu)\leq 1$ and is Lipchitz continuous in a suitable Banach space norm as shown in Section \ref{wellposedness}. 
The phase field is defined by 
\begin{align}
    \label{gammaplus}
    \gamma_+(t,\yy,\xx,\uu))=\overline{\mu}^{-1}\frac{\partial_{r^*}g^+(r^*(t,\yy,\xx,\uu))}{(S^*-P^*(t,\yy,\xx,\uu)))\sqrt{{|\yy-\xx|}/L}},  
\end{align}
for $\beta^+$ fixed and chosen in $0\leq \beta^+\leq 1$. and the plastic strain $P^*(\yy,\xx,t,\uu)$ is given by
\begin{align}
    \label{plastics}
    {P^*}(t,\yy,\xx,\uu)&=\beta_+\left(S^*(t,\yy,\xx,\uu)-\frac{\partial_{r^*} g^+(r^*(t,\yy,\xx,\uu))}{\overline{\mu}\sqrt{{|\yy-\xx|}/L}}\right)\nonumber\\
    &=S^\ast-\frac{\partial_{r^*} g^+(r^*(t,\yy,\xx,\uu))}{\overline{\mu}\gamma_+(t,\yy,\xx,\uu)\sqrt{{|\yy-\xx|}/L}}.
\end{align}
It follows from \eqref{gammaplus} that 
\begin{align}
    \label{check22}
    \overline{\mu}\gamma_+(t)\left(S^*(t,\yy,\xx,\uu)-P^*(t,\yy,\xx,\uu)\right)= \frac{\partial_{r} g^+(r^*(t,\yy,\xx,\uu))}{\sqrt{{|\yy-\xx|}/L}},
\end{align}

The elastic strain for bonds that can exhibit tensile failure and compressive failure is given by
\begin{align}
    \label{general elastic}
  E(\yy,\xx,\uu(t))=S(\yy,\xx,\uu(t))-(P^\ast(t,\yy,\xx,\uu)+  P_\ast(t,\yy,\xx,\uu)).
\end{align}
The extended constitutive law is of the form 
\begin{equation}\label{extendmugamma2}
\sigma(t,\yy,\xx,\uu):=\left\{\begin{array}{lc} \overline{\mu}\gamma_+(t)E(\yy,\xx,\uu(t)),&E(\yy,\xx,\uu(t))\geq 0,\\
0,&E(\yy,\xx,\uu(t))= 0,\\
\overline{\mu}\gamma_-(t)E(\yy,\xx,\uu(t)), & E(\yy,\xx,\uu(t))< 0.
\end{array}\right.
\end{equation}
The bond force per unit volume to elastic strain is described by
\begin{equation}
	\boldsymbol{f}^\epsilon(t,\yy,\xx,\uu)=\rho^\epsilon(\yy,\xx) \sigma(t,\yy,\xx,\uu)\boldsymbol{e},
	\label{Eqn.extendedconst}
\end{equation}
where $\sigma$ is defined by \eqref{extendmugamma2}.  
In Section \ref{wellposedness} we show that $\gamma_\pm(t,\yy,\xx,\uu)$, $P^\ast(t,\yy,\xx,\uu)$, and $P_\ast(t,\yy,\xx,\uu)$ are absolutely continuous in $t$ for almost every $(\yy,\xx)$.  Hence $\partial_t\gamma_\pm$ exists and $\gamma_\pm(t,\yy,\xx,\uu)=\int_0^t\partial_\tau\gamma_\pm(\tau,\yy,\xx,\uu)\,d\tau$, etc.

{\bf In summary the extended constitutive law features five properties:}
\begin{enumerate}
    \item There now two phase fields  $\gamma_\pm(t,\yy,\xx,\uu)$ one being a phase field in tension $\gamma_+(t,\yy,\xx,\uu)$ and the second  given by a phase field in compression $\gamma_{-}(t,\yy,\xx,\uu)$, with $0\leq \gamma_\pm \leq 1$.
    \item When the bond between $\yy$ and $\xx$ is in tension the compressive phase field $\gamma_-(t,\yy,\xx,\uu)$ is constant in $t$ and 
    when the bond is in compression the tensile phase field $\gamma_+(t,\yy,\xx,\uu)$ is constant in $t$.
    \item When the bond between $\yy$ and $\xx$ is under tension then $S_\ast(t,\yy,\xx,\uu)$ is constant in $t$ and when the bond is in compression $S^\ast (t,\yy,\xx,\uu)$ is constant in $t$.
    \item When the bond between $\yy$ and $\xx$ is under tension  $P_\ast(\yy,\xx,\uu)$ is constant in $t$ and when the bond is under compression  $P^\ast(t,\yy,\xx,t,\uu)$ is constant in $t$.
    \item $P^\ast(t,\yy,\xx,\uu)=0$ for $S^\ast (t,\yy,\xx,\uu)\leq S_t^Y$  and is nonzero and positive for $S^\ast (\yy,\xx,\uu)\geq S_t^Y$ and $P_\ast(t,\yy,\xx,\uu)=0$ for $S_\ast (t,\yy,\xx,\uu)\geq S_c^Y$ and is nonzero and negative for $S_\ast (t,\yy,\xx,\uu)\leq S_c^Y$
\end{enumerate}
For this case, the initial boundary value problem for fracture evolution is to
find a displacement $\uu(t,\xx)$ on $\Omega^*$ that satisfies 
\begin{align}\label{eq: prescrib boundary displacement3}
\uu(t,\xx)=\UU(t,\xx)& \hbox{ for }\xx \hbox{ in } \Omega^\epsilon_D,
\end{align}
with he nonlocal force density $\LL^\epsilon[\uu](t,\xx)$ defined for all points $\xx$ in $\Omega$ is  given by
\begin{align}
      \label{eq: force3}
 \LL^\epsilon [\uu](t,\xx) = -\int_{\Omega^*} \boldsymbol{f}^\epsilon(t,\yy,\xx,\uu) \,d\yy.
\end{align}
and  satisfies the evolution equation 
\begin{align}\label{eq: linearmomentumbal3}
\rho\ddot{\uu}(t,\xx)+ \LL^\epsilon [\uu](t,\xx) =\bb(t,\xx), \hbox{for $\xx$ in
$\Omega$ and $0<t<T$}
\end{align}
and satisfies the initial conditions on the displacement and velocity given by
\begin{align}
\uu(0,\xx)=\uu_0(\xx),\nonumber\\
\dot{\uu}(0,\xx)=\vv_0(\xx).\label{initialconditions3}
\end{align}
To illustrate ideas, we fix all bonds crossing the interface between $\Omega$ and $\Omega^\epsilon_D$ to be purely elastic throughout the evolution and have stiffness $\overline{\mu}$ as well as all bonds between points in $\Omega^\epsilon_D$.
In summary, the  general initial boundary problem with two phase fields suitable for bonds that evolve in both compression and tension is  given by  the evolution equation \eqref{eq: linearmomentumbal3}, initial conditions \eqref{initialconditions3}, and displacement boundary conditions \eqref{eq: prescrib boundary displacement3},   but now governed by the general constitutive law \eqref{extendmugamma2}.

We conclude this section noting that in the peridynamic taxonomy the blended
model can be classified as a history dependent ordinary state based peridynmic material model. To see this we have the symmetries $\gamma_\pm(t,\yy,\xx,\uu)=\gamma_\pm(t,\xx,\yy,\uu)$ and $E(\yy,\xx,\uu(t))=E(\xx,\yy,\uu(t))$ and we write
the bond force \eqref{Eqn.extendedconst} as
\begin{align}
	\boldsymbol{f}^\epsilon(t,\yy,\xx,\uu)&=\rho^\epsilon(\yy,\xx) \sigma(t,\yy,\xx,\uu)\boldsymbol{e}\nonumber\\
    &=T[\xx]\langle\yy-\xx\rangle-T[\yy]\langle\xx-\yy\rangle,
	\label{Eqn.state}
\end{align}
where the bond force state is history dependent and given by
\begin{equation}\label{eqn. bondforcestate}
T[\xx]\langle\yy-\xx\rangle=\frac{(\yy-\xx)}{2|\yy-\xx|}\rho^\epsilon(\yy,\xx)\left\{\begin{array}{lc} \overline{\mu}\gamma_+(t)E(\yy,\xx,\uu(t)),&E(\yy,\xx,\uu(t))\geq 0,\\
0,&E(\yy,\xx,\uu(t))= 0,\\
\overline{\mu}\gamma_-(t)E(\yy,\xx,\uu(t)), & E(\yy,\xx,\uu(t))< 0.
\end{array}\right.
\end{equation}
We can apply formulas (6.3) and (6.4) of \cite{sillinglehoucq2010} to the force state now defined by \eqref{Eqn.state} and \eqref{eqn. bondforcestate} to get the stress $\tau(\xx)$ at $\xx$. Or changing variables $\yy=\xx+\epsilon \boldsymbol{\xi}$ with $\boldsymbol{\xi}$ on the unit ball $\mathcal{H}_1(0)$ and using the result of  \cite{LiLiLaiLiu}
we can formally write the associated stress  $\tau(\xx)$ as
\begin{align}
\label{stress}
\tau(\xx)=-\frac{1}{2}\int_{{\mathcal{H}}_1(0)}\int_0^1{\boldsymbol{f}}^\epsilon(t,\xx+\alpha\boldsymbol{\xi},\xx-(1-\alpha)\boldsymbol{\xi},\uu)\otimes\boldsymbol{\xi}\,d\alpha\,d\boldsymbol{\xi}.
\end{align}
So one recovers the stress field a-posteriori from the displacement field solution of the IBVP.



\begin{figure}
    \centering
   
    \begin{subfigure}{.45\linewidth}
        \begin{tikzpicture}[scale=0.7]

\draw[thin] (-4,0) -- (5,0); 
\node[right] at (5,0){$S$};
\draw[thin] (0,-4) -- (0,3);    
\node[left] at (1.55,3) {$\sigma$};   


\draw[dark green, thick,
  postaction={
    decorate,
    decoration={
      markings,
      mark=at position 0.5 with {\arrow{>}}
    }
  }
] (0,0)--(1.5,2.5);

\node[left] at (0.2,0.2) {$A$}; 

\draw[black, dashed,
  postaction={
    decorate,
    decoration={
      markings,
      mark=at position 0.5 with {\arrow{>}}
    }
  }
] (1.5,2.5) -- (4.5,0.75);
\draw[black, dashed,
  postaction={
    decorate,
    decoration={
      markings,
      mark=at position 0.5 with {\arrow{>}}
    }
  }
]  (1.5,2.5)--(4.5,0.75); 
\draw[dark green, thick,
  postaction={
    decorate,
    decoration={
      markings,
      mark=at position 0.5 with {\arrow{>}}
    }
  }
]  (4.5,0.75)--(2.5,0); 
\draw[blue, thick,
  postaction={
    decorate,
    decoration={
      markings,
      mark=at position 0.5 with {\arrow{>}}
    }
  }
]  (2.5,0)--(1,-2.5);
\node[left] at (2.7,0.2) {$B$}; 

\draw[black, dashed,
  postaction={
    decorate,
    decoration={
      markings,
      mark=at position 0.5 with {\arrow{>}}
    }
  }
]  (1,-2.5) to [out = 235, in = -20] (-3.0,-1.33);

\draw[blue, thick,
  postaction={
    decorate,
    decoration={
      markings,
      mark=at position 0.5 with {\arrow{>}}
    }
  }
]  (-3.0,-1.333)--(-1,0);
\node[left] at (-0.8,0.2) {$C$}; 
\end{tikzpicture}
        \caption{}
		\label{tensionCompressiveConstitutiveLaw}
    \end{subfigure}
     \hskip2em
\begin{subfigure}{.45\linewidth}
        \begin{tikzpicture}[scale=0.7]

\draw[thin] (-4,0) -- (5,0);   
\node[right] at (5,0) {$E$};
\draw[thin] (0,-4) -- (0,3);    
\node[right] at (0.15,3) {$\sigma$};   

\draw[dark green, thick,
  postaction={
    decorate,
    decoration={
      markings,
      mark=at position 0.5 with {\arrow{>}}
    }
  }
] (0,0)--(1.5,2.5);

        \node[right] at (0.5,0.85) {\color{dark green}{$\bar{\mu}$}}; 

\draw[dashed, thick,
  postaction={
    decorate,
    decoration={
      markings,
      mark=at position 0.5 with {\arrow{<}}
    }
  }
  ] (2.0,0.75) -- (1.5,2.5);



\draw[dark green, thick,
  postaction={
    decorate,
    decoration={
      markings,
      mark=at position 0.5 with {\arrow{>}}
    }
  }
]  (2.0,0.75)--(0,0);

        \node[right] at (1.25,0.35) {\color{dark green}{$\gamma_+\bar{\mu}$}}; 
        \node[right] at (-0.75,-1.35) {\color{blue}{$\bar{\mu}$}}; 
\draw[dashed, thick,
  postaction={
    decorate,
    decoration={
      markings,
      mark=at position 0.5 with {\arrow{>}}
    }
  }
]  (-1.5,-2.5) to [out = 235, in = -20] (-2,-1.33);

\draw[blue, thick,
  postaction={
    decorate,
    decoration={
      markings,
      mark=at position 0.5 with {\arrow{>}}
    }
  }
]  (0,0)--(-1.5,-2.5);


\draw[blue, thick,
  postaction={
    decorate,
    decoration={
      markings,
      mark=at position 0.5 with {\arrow{>}}
    }
  }
]  (-2.00,-1.333)--(0,0);
        \node[right] at (-1.8,-0.38) {\color{blue}{$\gamma_-\bar{\mu}$}}; 
\end{tikzpicture}
		  \caption{}
		  \label{tension-compresssion-const-law}
    \end{subfigure}
   
    \caption{(a) A bond that degrades in tension and compression undergoing  stress-strain cycling plotted in $(S,\sigma)$  coordinates to illustrate the evolution of the sum of irreversible tensile and compressive strain $P^\ast(t,\yy,\xx,\uu)+P_\ast(t,\yy,\xx,\uu)$ At the beginning of this cycle $A$ there is no  plastic strain, after tension loading there is tensile plastic stress $B$ after compressive loading  the sum of the plastic stress under compression and tension is given by $C$. (b) A bond that degrades in tension and compression undergoing  stress-strain cycling now plotted in $(E,\sigma)$  coordinates to show stiffness degradation under tension and compression. Here the cycle starts with strain loading  and then unloads  then reverses and is loaded in compression  and again unloads. Both figures correspond to the same loading cycle with bilinear tensile loading followed by linear-exponential compression loading.}
    \label{Compressive constitutive law}
\end{figure}
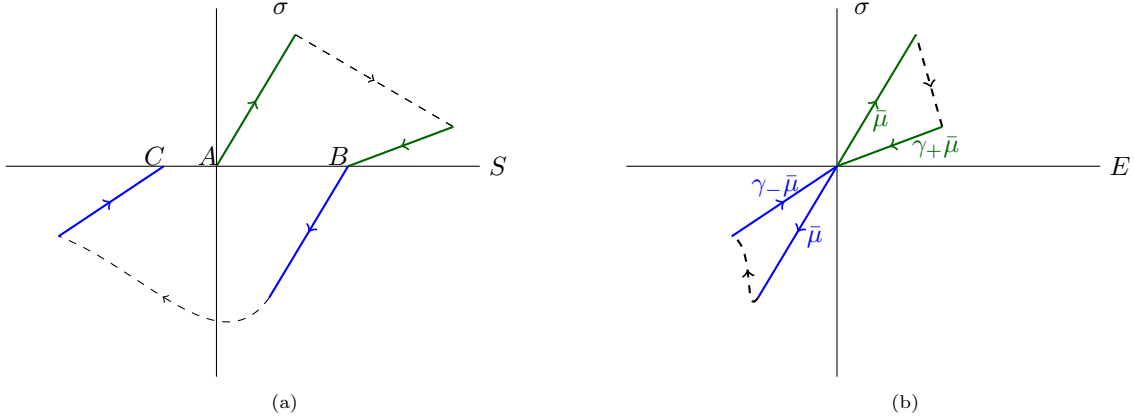


\section{Existence and uniqueness of displacement and damage evolution.}
\label{wellposedness}

In this section, we state and prove the existence and uniqueness of solution for the fracture evolution subject to body loading and boundary displacement and general cyclic constitutive law given by the vector phase field described in Section \ref{Generalization}. To expedite the presentation we sketch the parts of proof already covered in \cite{lipton2024energy} and \cite{coskundamerchelilipton} and elaborate the proof steps related to the general cyclic constitutive law.

We recall the space of all bounded measurable functions $L^\infty(\Omega;\mathbb{R}^d)$.  To expedite the presentation, we denote $ L^\infty(\Omega;\mathbb{R}^d)$ by $X$. The norm on $C^2([0,T];X)$ is given by \eqref{norms}. 
\begin{align}\label{norms}
\Vert \uu(t,\xx)\Vert_{C^2([0,T]; X)}=\sup_{0\leq t\leq T}\{ \sum_{i=0}^2||\partial_t^i\uu(t,\xx)||_{X}\}.
\end{align}
To simplify the notation, we denote the space $C([0,T];X)$ as $CX$ with the corresponding norm $\Vert\cdot \Vert_{CX}$. Similarly, we write $C^2([0,T];X)$ as $C^2X$.  To characterize all admissable boundary loads we introduce the Banach space $X^*=L^\infty(\Omega^*,\mathbb{R}^d)$, and the Banach space $C^1([0,T];X^*)$. The imposed boundary displacement $\UU(t,\xx)$ belongs to $C^1([0,T];X^*)$ and is zero for $\xx$ in $\Omega$ and is a prescribed nonzero function $\UU(t,\xx)$ on $\Omega_D^\epsilon$. 

\begin{theorem}[{\bf Existence and Uniqueness of Solution of the Displacement and Force Controlled Fracture Evolution}]
\label{exist2}
 The initial boundary value problem  given by \eqref{extendmugamma2}, \eqref{eq: prescrib boundary displacement3}, \eqref{eq: linearmomentumbal3}, and \eqref{initialconditions3}  with initial data in $X^*$, $\bb(t,\xx)$ belonging to $CX$, and imposed boundary displacement $\UU(t,\xx)$ in $C^1X^*$,
has a unique solution $\uu(t,\xx)$ belonging to $C^1X^*$ and $C^2X$ with two strong derivatives in time.
 \end{theorem}

The initial value problem delivers a unique displacement phase field pair: ${\uu^\epsilon(t), \gamma_\pm(t)}$. 
The key points are that the operator $\LL^\epsilon\uu$ is a map from $CX$ into itself, and it is Lipschitz continuous in $\uu$ with respect to the $CX$ norm. The existence and uniqueness of solution then follows from the Banach fixed-point theorem. 


We provide a sketch of the proof of Theorem \ref{exist2}. The theorem follows from the Lipchitz continuity of the stress field $\sigma(t,\yy,\xx,\uu)$   with respect to the displacement $\uu$ and this follows from a two step calculation. The first calculation similar to \cite{coskundamerchelilipton} delivers the point wise estimates
 \begin{align}\label{sast}
     |S^*(\yy,\xx,\uu,t) - S^*(\yy,\xx,\ww,t)| \leq \max_{0\leq\tau \leq t}|S(\yy,\xx,\uu(\tau)-\ww(\tau))| \leq \frac{C}{|\yy-\xx|} \Vert\uu-\ww\Vert_{CX},
 \end{align}    
 \begin{align}\label{spowerast}
     |S_*(\yy,\xx,\uu,t) - S_*(\yy,\xx,\ww,t)| \leq \max_{0\leq\tau \leq t}|S(\yy,\xx,\uu(\tau)-\ww(\tau))| \leq \frac{C}{|\yy-\xx|} \Vert\uu-\ww\Vert_{CX},
 \end{align}
 \begin{align}\label{past}
 |P^*(t,\yy,\xx,\uu)-P^*(t,\yy,\xx,\ww)| \leq C\max_{0\leq\tau \leq t}|S(\yy,\xx,\uu(\tau)-\ww(\tau))|\leq \frac{C}{|\yy-\xx|}\Vert\uu-\ww\Vert_{CX},
 \end{align}
 and
 \begin{align}\label{plowerast}
 |P_*(t,\yy,\xx,\uu)-P_*(t,\yy,\xx,\ww)| \leq C\max_{0\leq\tau \leq t}|S(\yy,\xx,\uu(\tau)-\ww(\tau))|\leq \frac{C}{|\yy-\xx|}\Vert\uu-\ww\Vert_{CX},
 \end{align}
where $C$ is a constant independent of $\uu$ and $\ww$. 
Hence
 \begin{align}\label{elowerast}
 |E(t,\yy,\xx,\uu)-E(t,\yy,\xx,\ww)| \leq C\max_{0\leq\tau \leq t}|S(\yy,\xx,\uu(\tau)-\ww(\tau))|\leq \frac{C}{|\yy-\xx|}\Vert\uu-\ww\Vert_{CX},
 \end{align}
 and
 \begin{align}\label{gammapm}
 |\gamma_\pm(t,\yy,\xx,\uu)-\gamma_\pm(t,\yy,\xx,\ww)| \leq C\max_{0\leq\tau \leq t}|S(\yy,\xx,\uu(\tau)-\ww(\tau))|\leq \frac{C}{|\yy-\xx|}\Vert\uu-\ww\Vert_{CX}.
 \end{align}
The second and final step is new and necessary for proving Lipchitz continuity of the stress field $\sigma(t,\yy,\xx,\uu)$  for general cyclic constitutive laws. This step accounts for both the tension and compression states of a bond associated with two independent displacement fields $\uu$ and $\ww$. With this in mind we demonstrate Lipchitz continuity for the four conditions on the stress free strain given by
\begin{align}
    \label{timestates}
    \begin{cases}
E(\yy,\xx,\uu(t))>0,E(\yy,\xx,\ww(t))>0,\\
E(\yy,\xx,\uu(t))<0,E(\yy,\xx,\ww(t))<0,\\
E(\yy,\xx,\uu(t))<0,E(\yy,\xx,\ww(t))>0,\\
E(\yy,\xx,\uu(t))>0,E(\yy,\xx,\ww(t))<0
\end{cases}
\end{align}
Point-wise Lipchitz estimates are shown for the first and third case listed above noting that the second and fourth case follow in the same way. For ease of exposition we let $\gamma_+(\sqrt{|\yy-\xx|/L}S^\ast(t,\yy,\xx,\uu))=\gamma_+(S^\ast(t,\uu))$ and $E(\yy,\xx,\uu(t))=E(\uu(t))$. For the first case we have
\begin{align}
    \label{firstcase}
   & |\sigma(t,\yy,\xx,\uu)-\sigma(t,\yy,\xx,\ww)|=|\overline{\mu}\gamma_+(S^\ast(t,\uu))E(\uu(t))-\overline{\mu}\gamma_+(S^\ast(t,\ww))E(\ww(t))|\nonumber \\
   &\leq|\overline{\mu}\gamma_+(S^\ast(t,\uu))E(\uu(t))-\overline{\mu}\gamma_+(S^\ast(t,\uu))E(\ww(t))|+|[\overline{\mu}\gamma_+(S^\ast(t,\uu))- \overline{\mu}\gamma_+(S^\ast(t,\ww))]E(\ww(t))|\nonumber\\
   &\leq\overline{\mu}|E(\uu(t))-E(\ww(t))|+\overline{\mu}(1+\beta_++\beta_-)(1+S_t^F+|S^F_c|)|\gamma_+(S^\ast(t,\uu))-\gamma_+(S^\ast(t,\ww))|\nonumber\\
   &\leq C \max_{0\leq\tau \leq t}|S(\yy,\xx,\uu(\tau)-\ww(\tau))|\leq \frac{C}{|\yy-\xx|} \Vert\uu-\ww\Vert_{CX}.
\end{align}
where the first term in the third inequality follows from $0\leq\gamma_+\leq1$, and the second term in the third inequality follow from bounds on the maximum strain $|E(\ww(t))|$ corresponding to nonzero post peak stress, the fourth inequality follows from \eqref{sast} and \eqref{past}  together with the Lipchitz continuity of $\gamma_+$ with respect to the variable $S^\ast$, see Figure \ref{ConvexConcaveC}.
For the third case we have 
\begin{align}
    \label{thirdcase}
   |\sigma(t,\yy,\xx,\uu)-\sigma(t,\yy,\xx,\ww)| &=|\overline{\mu}\gamma_-(S_\ast(t,\uu))E(\uu(t))-\overline{\mu}\gamma_+(S^\ast(t,\ww))E(\ww(t))|\nonumber \\
   &\leq|\overline{\mu}E(\uu(t))-\overline{\mu}E(\ww(t))|\leq \frac{C}{|\yy-\xx|} \Vert\uu-\ww\Vert_{CX}.
\end{align}
where the last inequality follows from \eqref{elowerast}.
So for all for cases given by \eqref{timestates} we conclude
\begin{align}
    \label{multicase}
   |\sigma(t,\yy,\xx,\uu)-\sigma(t,\yy,\xx,\ww)| & \leq \frac{C}{|\yy-\xx|} \Vert\uu-\ww\Vert_{CX},
\end{align}
and it now follows that
$$\Vert \mathcal{L}^\epsilon[\uu]-\mathcal{L}^\epsilon[\ww]\Vert_{CX}\leq C \Vert\uu-\ww\Vert_{CX}$$
With this result in hand 
Theorem \ref{exist2} is established using the Banach fixed point theorem  as in   \cite{lipton2024energy}, \cite{coskundamerchelilipton}, \cite{EmmrichPhust}.

We get the absolute continuity in time for the phase fields and plastic strains as an immediate consequence of the existence of solution in $C^2X$. This is summarized in
\begin{lemma}
    The solution $\uu$ of the IBVP given by Theorem \ref{exist2} belongs to $C^2X$, hence, the mapping 
    \begin{align}
    \label{abscont}
    t\mapsto (\gamma_+(t,\yy,\xx,\uu), \gamma_-(t,\yy,\xx,\uu), P^\ast(t,\yy,\xx,\uu), P_\ast(t,\yy,\xx,\uu)):\mathbb{R}\rightarrow\mathbb{R}^4
    \end{align}
    is absolutely continuous for a.e. $(\yy,\xx)$ in $\Omega\times\Omega$.
    \end{lemma}
To establish this lemma note as in \cite{lipton2024energy} that if $\uu$ is the solution of the IBVP then it belongs to $C^2X$ so $S^\ast(t,\yy,\xx,\uu)$ and $S_\ast(t,\yy,\xx,\uu)$ are Lipchitz continuous in time. The Lemma follows noting further that $\gamma_+(S^\ast)$, $\gamma_-(S_\ast)$, $\partial_rg^+(S^\ast)$ and $\partial_rg^-(S_\ast)$ are Lipchitz continuous in their arguments.




\section{Power balance for fracture evolutions with damage and plasticity}
\label{powerbalanceprocesszone2}

In this section, we establish the power balance  for fracture evolutions with body force and boundary displacement loadings. Here, the boundary displacement $\UU(t,\xx)$  and body force $\bb(t,\xx)$ are prescribed.  Multiplying the linear momentum equation given in \eqref{eq: linearmomentumbal2} with $\dot{\uu}(t,\xx)$ and integrating over the domain $\Omega$ results in 
 \begin{equation}
	\int_{\Omega} \rho \ddot{\uu}(\xx, t) \cdot \dot{\uu}(\xx, t) \,d\xx + 
	\int_{\Omega} \LL^\epsilon [\uu](t,\xx) \cdot \dot{\uu}(\xx, t) \,d\xx = 
	\int_{\Omega} \bb(t,\xx) \cdot \dot{\uu}(\xx, t) \,d\xx.
\label{eq: linearmomentumbalInt2}
\end{equation}
We apply the identity $\chi_{\Omega^*}=\chi_\Omega+\chi_{\Omega_D^\epsilon}$ integrate by parts noting that $S^*(t,\yy,\xx,\uu)$,  $S_*(t,\yy,\xx,\uu)$, $P_*(t,\xx,\yy,\uu)$ and, $P_*(t,\xx,\yy,\uu)$  are invariant under swapping  $\yy$ and $\xx$ and a careful calculation  gives
\begin{align}\label{eq:prebalance2}
&\dot{\mathcal{K}}(t)+\frac{1}{2}\int_\Omega\int_\Omega\,|\yy-\xx|\rho^{\epsilon}(\yy,\xx)\overline{\mu}\gamma_+(t,\yy,\xx,\uu) E(\yy,\xx,\uu(t)) E(\yy,\xx,\dot\uu(t))\,d\yy\,d\xx\nonumber\\
&\frac{1}{2}\int_\Omega\int_\Omega\,|\yy-\xx|\rho^{\epsilon}(\yy,\xx)\overline{\mu}\gamma_-(t,\yy,\xx,\uu) E(\yy,\xx,\uu(t)) E(\yy,\xx,\dot\uu(t))\,d\yy\,d\xx\nonumber\\
&+\frac{1}{2}\int_\Omega\int_\Omega\,|\yy-\xx|\rho^{\epsilon}(\yy,\xx)\overline{\mu}\gamma_+(t,\yy,\xx,\uu) E(\yy,\xx,\uu(t)) (\dot{P}^*(t,\yy,\xx,\uu)+\dot{P}_\ast(t,\yy,\xx,\uu))\,d\yy\,d\xx\nonumber\\
&+\frac{1}{2}\int_\Omega\int_\Omega\,|\yy-\xx|\rho^{\epsilon}(\yy,\xx)\overline{\mu}\gamma_-(t,\yy,\xx,\uu) E(\yy,\xx,\uu(t)) (\dot{P}^*(t,\yy,\xx,\uu)+\dot{P}_\ast(t,\yy,\xx,\uu))\,d\yy\,d\xx\nonumber\\
& + \dot{\mathcal{I}}^\epsilon(t)= \dot{\mathcal{R}}^\epsilon(t)+\int_{\Omega} \bb(t,\xx) \cdot \dot{\uu}(\xx, t) \,d\xx,
\end{align}
where
\begin{align}
\label{kenitic}
    \mathcal{K}(t)=\int_\Omega\frac{\rho|\dot\uu(t)|^2}{2}\,d\xx,
\end{align}
is the Kinetic energy and recalling all bonds connecting the displacement loading region are elastic and undamaged gives that the power is related to the change in elastic energy density there, i.e.,
\begin{align}
    \label{interaction}
    \dot{\mathcal{I}}^\epsilon(t)&=\frac{1}{2}\int_\Omega\int_{\Omega_D^\epsilon}|\yy-\xx|\frac{J^\epsilon(|\yy-\xx|)}{\epsilon V_d^\epsilon}\partial_t\left(\bar{\mu}\frac{E^2(\yy,\xx,\uu(t))}{2}\right)d\yy d\xx\nonumber\\
    &+\frac{1}{2}\int_{\Omega_D^\epsilon}\int_{\Omega}|\yy-\xx|\frac{J^\epsilon(|\yy-\xx|)}{\epsilon V_d^\epsilon}\partial_t\left(\bar{\mu}\frac{E^2(\yy,\xx,\uu(t))}{2}\right)d\yy d\xx
\end{align}
and on the displacement loading zone
\begin{align}
    \label{forceflux}
    \dot{\mathcal{R}}^\epsilon(t)=\int_{\Omega_D^\epsilon}\int_{\Omega}\frac{J^\epsilon(|\yy-\xx|)}{\epsilon V_d^\epsilon}\bar{\mu}E(\yy,\xx,\uu(t))\ee_{\xx-\yy}\cdot\dot{\UU}(t,\xx)d\yy d\xx.
\end{align}

To obtain power balance, we rewrite the second term  in \eqref{eq:prebalance2} using the product rule for the time derivative
to get
\begin{align}\label{eq:2ndterm}
\frac{1}{2}\int_\Omega\int_\Omega\,|\yy-\xx|\rho^{\epsilon}(\yy,\xx)\partial_t\left(\overline{\mu}\gamma_+(t)\frac{E(\yy,\xx,\uu(t))^2}{2}\right)-\partial_t(\overline{\mu}\gamma_+(t))\frac{E(\yy,\xx,\uu(t))^2}{2}\,d\yy\,d\xx,
\end{align}
The third term is rewritten in a similar way and upon substitution and rearrangement  in \eqref{eq:prebalance2} we get
\begin{align}\label{eq:prebalance3}
&\dot{\mathcal{K}}(t)+\frac{1}{2}\int_\Omega\int_\Omega\,|\yy-\xx|\rho^{\epsilon}(\yy,\xx))\partial_t\left(\overline{\mu}\gamma_+(t)\frac{E(\yy,\xx,\uu(t))^2}{2}\right)\,d\yy\,d\xx\nonumber\\
&+\frac{1}{2}\int_\Omega\int_\Omega\,|\yy-\xx|\rho^{\epsilon}(\yy,\xx)\overline{\mu}\gamma_+(t) E(\yy,\xx,\uu(t)) (\dot{P}^*(t,\yy,\xx,\uu)+\dot{P}_\ast(t,\yy,\xx,\uu))-\partial_t(\overline{\mu}\gamma_+(t))\frac{E(\yy,\xx,\uu(t))^2}{2}\,d\yy\,d\xx\nonumber\\
&+\frac{1}{2}\int_\Omega\int_\Omega\,|\yy-\xx|\rho^{\epsilon}(\yy,\xx))\partial_t\left(\overline{\mu}\gamma_-(t)\frac{E(\yy,\xx,\uu(t))^2}{2}\right)\,d\yy\,d\xx\nonumber\\
&+\frac{1}{2}\int_\Omega\int_\Omega\,|\yy-\xx|\rho^{\epsilon}(\yy,\xx)\overline{\mu}\gamma_-(t) E(\yy,\xx,\uu(t)) (\dot{P}^*(t,\yy,\xx,\uu)+\dot{P}_\ast(t,\yy,\xx,\uu))-\partial_t(\overline{\mu}\gamma_-(t))\frac{E(\yy,\xx,\uu(t))^2}{2}\,d\yy\,d\xx\nonumber\\
& + \dot{\mathcal{I}}^\epsilon(t)= \dot{\mathcal{R}}^\epsilon(t)+\int_{\Omega} \bb(t,\xx) \cdot \dot{\uu}(\xx, t) \,d\xx,
\end{align}

Next we examine the integrand of the third  integral in \eqref{eq:prebalance3}.
We begin by noting  that $\partial_t\gamma_+(t,\yy,\xx,\uu)<0$, $\dot{P}_*(\yy,\xx,\uu)=0$, $\dot{P}^*(\yy,\xx,\uu)\not=0$ and positive only when  $S(\yy,\xx,\uu(t))=S^*(t,\yy,\xx,\uu)$ and $\partial_t\gamma_-(t,\yy,\xx,\uu)<0$, $\dot{P}^*(\yy,\xx,\uu=0$, $\dot{P}_*(\yy,\xx,\uu)\not=0$ and negative only when  $S(\yy,\xx,\uu(t))=S_*(t,\yy,\xx,\uu)$. Suppose first that  $S(\yy,\xx,\uu(t))=S^*(t,\yy,\xx,\uu)$ then applying formula \eqref{check22} we find that
\begin{align}
\label{changeP}
&\overline{\mu}\gamma_+(t) E(\yy,\xx,\uu(t)) \dot{P}^*(t,\yy,\xx,\uu)=\frac{\partial_{r}g^+(r^*(t))}{\sqrt{|\yy-\xx|/L}}\dot{P}^*(t,\yy,\xx,\uu)-\overline{\mu}\gamma_+(t)\dot{P}^{^\ast}(t,\yy,\xx,\uu)P_\ast(t,\yy,\xx,\uu),\nonumber\\
&=\frac{L}{|\yy-\xx|}\left[\partial_tg^+(r^*(t))-\partial_r g^+(r^*(t))\partial_t\left(\frac{\partial_r g^+(r^*(t))}{\overline{\mu}\gamma_+(t)}\right)\right]-\overline{\mu}\gamma_+(t)\dot{P}^{^\ast}(t,\yy,\xx,\uu)P_\ast(t,\yy,\xx,\uu).
\end{align}
Note that when $\dot\gamma_+(t,\yy,\xx,\uu)\not=0$ we have
\begin{align}
\label{ident2}
-\partial_t(\overline{\mu}\gamma_+(t))\frac{(E(\yy,\xx,\uu(t))^2}{2}&=-\frac{L}{|\yy-\xx|}\frac{\partial_t\gamma_+(t)(\partial_{r}g^+(r^*(t)))^2}{2\overline{\mu}\gamma_+^2(t)}\nonumber\\
&+\overline{\mu}\partial_t\gamma_+\left(\frac{\partial_rg^+(r^\ast(t))}{\overline{\mu}\gamma_+(t)\sqrt{|\yy-\xx|/L}}P_\ast(t,\yy,\xx,\uu)\right)-\frac{1}{2}\overline{\mu}\partial_t\gamma_+(t)P^2_\ast(t,\yy,\xx,\uu).
\end{align}
where each term on the right hand side of \eqref{ident2} is positive and we get
\begin{align}
\label{diss2nd}
&\overline{\mu}\gamma_+(t) E(\yy,\xx,\uu(t)) \dot{P}^*(t,\yy,\xx,\uu)-\partial_t(\overline{\mu}\gamma_+(t))\frac{E(\yy,\xx,\uu(t))^2}{2}\nonumber \\
&=\frac{L}{|\yy-\xx|}\partial_t\left[g^+(r^*(t))-\frac{(\partial_{r^*}g^+(r^*(t)))^2}{2\overline{\mu}\gamma_+(t)}\right]-\overline{\mu}\gamma_+(t)\dot{P}^{^\ast}(t,\yy,\xx,\uu)P_\ast(t,\yy,\xx,\uu)\nonumber\\
&+\overline{\mu}\partial_t\gamma_+\left(\frac{\partial_rg^+(r^\ast(t))}{\overline{\mu}\gamma_+(t)\sqrt{|\yy-\xx|/L}}P_\ast(t,\yy,\xx,\uu)\right)-\frac{1}{2}\overline{\mu}\partial_t\gamma_+(t)P^2_\ast(t,\yy,\xx,\uu),
\end{align}
 where all terms on the right hand side of \eqref{diss2nd} are positive.

For  $S(\yy,\xx,\uu(t))=S_*(t,\yy,\xx,\uu)$ we see in the same way that 
\begin{align}
\label{diss3rd}
&\overline{\mu}\gamma_-(t) E(\yy,\xx,\uu(t)) \dot{P}_*(t,\yy,\xx,\uu)-\partial_t(\overline{\mu}\gamma_-(t))\frac{E(\yy,\xx,\uu(t))^2}{2}\nonumber \\
&=\frac{L}{|\yy-\xx|}\partial_t\left[g_-(r_*(t))-\frac{(\partial_{r_*}g^-(r_*(t)))^2}{2\overline{\mu}\gamma_-(t)}\right]-\overline{\mu}\gamma_-(t)\dot{P}_\ast(t,\yy,\xx,\uu)P^\ast(t,\yy,\xx,\uu)\nonumber\\
&+\overline{\mu}\partial_t\gamma_-(t)\left(\frac{\partial_rg^-(r_\ast(t))}{\overline{\mu}\gamma_-(t)\sqrt{|\yy-\xx|/L}}P^\ast(t,\yy,\xx,\uu)\right)-\frac{1}{2}\overline{\mu}\partial_t\gamma_-(t)(P^\ast(t,\yy,\xx,\uu))^2,
\end{align}
where all terms on the right hand side of \eqref{diss3rd} are positive.
Labeling the second and fourth integrals of \eqref{eq:prebalance3} as $\dot{\mathcal E}(t)$ and  third and fifth integrals of \eqref{eq:prebalance3} $\dot{\mathcal{D}}(t)$ respectively we get that the power related to the elasticity is given by
\begin{align}
\label{elasticpower}
\dot{\mathcal E}=\frac{1}{2}\int_\Omega\int_\Omega\,|\yy-\xx|\rho^{*,\epsilon}(\yy,\xx))\left[\partial_t\left(\overline{\mu}\gamma_+(t)\frac{E(\yy,\xx,\uu(t))^2}{2}\right)+\partial_t\left(\overline{\mu}\gamma_-(t)\frac{E(\yy,\xx,\uu(t))^2}{2}\right)\right]\,d\yy\,d\xx+ \dot{\mathcal{I}}^\epsilon(t)
\end{align}
and that  the dissipation  rate is given by
\begin{align}
\label{damagepower}
&\dot{\mathcal{D}}=\frac{L}{2}\int_\Omega\int_\Omega\,\rho^{\epsilon}(\yy,\xx)\partial_t\left[g^+(r^*(t))-\frac{(\partial_{r^*}g^+(r^*(t)))^2}{2\overline{\mu}\gamma_+(t)}\right]\, d\yy d\xx\nonumber \\
&-\frac{1}{2}\int_\Omega\int_\Omega\,|\yy-\xx|\rho^{\epsilon}(\yy,\xx)\overline{\mu}\gamma_+(t)\dot{P}^{^\ast}(t,\yy,\xx,\uu)P_\ast(t,\yy,\xx,\uu)\,d\yy\,d\xx\nonumber \\
&+ \frac{1}{2}\int_\Omega\int_\Omega\,|\yy-\xx|\rho^{\epsilon}(\yy,\xx)\overline{\mu}\partial_t\gamma_+(t)\left(\frac{\partial_rg^+(r^\ast(t))}{\overline{\mu}\gamma_+(t)\sqrt{|\yy-\xx|/L}}P_\ast(t,\yy,\xx,\uu)\right)-\frac{1}{2}\overline{\mu}\partial_t\gamma_+(t)(P_\ast(t,\yy,\xx,\uu))^2  \,d\yy\,d\xx \nonumber\\
&+\frac{L}{2}\int_\Omega\int_\Omega\,\rho^{\epsilon}(\yy,\xx)\partial_t\left[g^-(r_*(t))-\frac{(\partial_{r^*}g^-(r_*(t)))^2}{2\overline{\mu}\gamma_-(t)}\right]\, d\yy d\xx\nonumber \\
&-\frac{1}{2}\int_\Omega\int_\Omega\,|\yy-\xx|\rho^{\epsilon}(\yy,\xx) \overline{\mu}\gamma_-(t)\dot{P}_\ast(t,\yy,\xx,\uu)P^\ast(t,\yy,\xx,\uu)\,d\yy\,d\xx\nonumber\\
&+ \frac{1}{2}\int_\Omega\int_\Omega\,|\yy-\xx|\rho^{\epsilon}(\yy,\xx)\overline{\mu}\partial_t\gamma_-(t)\left(\frac{\partial_rg^-(r_\ast(t))}{\overline{\mu}\gamma_-(t)\sqrt{|\yy-\xx|/L}}P^\ast(t,\yy,\xx,\uu)\right)-\frac{1}{2}\overline{\mu}\partial_t\gamma_-(t)(P^\ast(t,\yy,\xx,\uu))^2  \,d\yy\,d\xx.
\end{align}
One readily checks  that the integrand in  first and fourth term of \eqref{damagepower} is positive since first time derivatives are increasing and  the positivity of the second, third, fifth and sixth terms term follow from the properties of $\gamma_\pm(t)$, $\partial_t\gamma_\pm(t)$, ${P}^\ast$, $\dot{P}^\ast$, ${P}_\ast$, and $\dot{P}_\ast$, hence the dissipation rate $\dot{\mathcal{D}}$ is either zero or positive in accord with thermodynamics.



On collecting results we obtain

\noindent{\bf Power balance for for fracture in quasi brittle materials with damage and plasticity}
\begin{align}\label{eq:rateenergybalance4}
\dot{\mathcal{K}}(t) + \dot{\mathcal{E}}^\epsilon(t) + \dot{\mathcal{D}}^\epsilon(t)=\dot{\mathcal{R}}(t)+\int_\Omega\,\bb(t)\cdot\dot\uu(t)\,d\xx.
\end{align}
Since $\dot{\mathcal{D}}^\epsilon(t)\geq 0$, these observations are summarized in the following Lemma.
\begin{lemma}[{\bf Growth of the dissipation for fracture in quasi-brittle materials}]
\label{damagezonegrowth4}
\begin{align}\label{eq:processzonepowerbalance4}
\dot{\mathcal{D}}^\epsilon(t) =\int_\Omega\,\bb(t)\cdot\dot\uu(t)\,d\xx+{\mathcal{R}}^\epsilon(t)-\dot{\mathcal{K}}(t) - \dot{\mathcal{E}}^\epsilon(t)  \geq 0.
\end{align}
\end{lemma}

The power balance is used to identify a condition when energy dissipation occurs.
\begin{corollary}[{\bf  Condition for energy dissipation in fracture evolution for quasi brittle materials}]
\label{damagecondition2}
{\em If the rate of energy put into the system exceeds the material's capacity to generate kinetic and elastic energy through elastic strain and velocity, then energy dissipation occurs  through damage manifested as softening elastic moduli and non-recoverable strain.}
\end{corollary}
\noindent Note here that \eqref{eq:processzonepowerbalance4} is consistent with thermodynamics and follows from the initial value problem directly through the equations of motion \eqref{eq: linearmomentumbal3} and the Dirichlet boundary conditions.

In summary, power balance for quasi-brittle crack growth shows that
$\dot{\mathcal{D}}>0$ only where damage is occurring and zero elsewhere.   If a crack exists, the strain is greatest in the neighborhood of its tips  and the region where $\dot{\mathcal{D}}>0$ corresponds to the process zone in front of a crack.

\section{Energy balance satisfied by fracture  evolutions.}
\label{energybalanceprocesszone2}

To obtain an  explicit formula for the elastic energy and dissipation energy, one exchanges time and space integrals in $\int_0^t\dot{\mathcal{E}}(\tau) d\tau$ and $\int_0^t\dot{\mathcal{D}}(\tau)\,d\tau$. 
Exchanging time and space integration and using \eqref{elasticpower} we get
\begin{align}
   \label{Antiderivativeedot}
    {\mathcal{E}}(t)&=\frac{1}{2}\int_\Omega\int_\Omega|\yy-\xx| \rho^{\epsilon}(\yy,\xx)\left(\overline{\mu}\gamma_+(t,\yy,\xx,\uu) \frac{E^2(\yy,\xx,\uu(t))}{2}\right)\,d\yy\,d\xx\nonumber  \\
    &=\frac{1}{2}\int_\Omega\int_\Omega|\yy-\xx| \rho^{\epsilon}(\yy,\xx)\left(\overline{\mu}\gamma_-(t,\yy,\xx,\uu) \frac{E^2(\yy,\xx,\uu(t)}{2}\right)\,d\yy\,d\xx\nonumber  \\
    &+{\mathcal{I}}^\epsilon(t).
\end{align}
and
\begin{align}
   \label{Initialenergyelastic}
  {\mathcal{E}}(0)=
&  \frac{1}{2}\int_\Omega\int_\Omega|\yy-\xx| \rho^{*,\epsilon}(\yy,\xx)\chi_{EZ^\epsilon(0)}\left(\overline{\mu} \frac{E^2(\yy,\xx,\uu(t)}{2}\right)\,d\yy\,d\xx + {\mathcal{I}}^\epsilon(0).
\end{align}

Recalling that $\mathcal{D}(0)=0$, exchanging space and time integration and collecting results shows the damage energy expended from $0$ to $t$ is given by the non-negative quantity

\begin{align}
\label{damageenergy}
&{\mathcal{D}}(t)=\frac{L}{2}\int_\Omega\int_\Omega\,\rho^{\epsilon}(\yy,\xx)\left[g^+(r^*(t))-\frac{(\partial_{r^*}g^+(r^*(t)))^2}{2\overline{\mu}\gamma_+(t)}\right]\, d\yy d\xx\nonumber \\
&-\frac{1}{2}\int_\Omega\int_\Omega\int_0^t\,|\yy-\xx|\rho^{\epsilon}(\yy,\xx)\,\overline{\mu}\gamma_+(\tau)\dot{P}^{^\ast}(\tau,\yy,\xx,\uu)P_\ast(\tau,\yy,\xx,\uu)\,d\tau\,d\yy\,d\xx\nonumber \\
&+ \frac{1}{2}\int_\Omega\int_\Omega\int_0^t\,|\yy-\xx|\rho^{\epsilon}(\yy,\xx)\overline{\mu}\partial_t\gamma_+(\tau)\left(\frac{\partial_rg^+(r^\ast(\tau))}{\overline{\mu}\gamma_+(\tau)\sqrt{|\yy-\xx|/L}}P_\ast(\tau,\yy,\xx,\uu)\right)-\frac{1}{2}\overline{\mu}\partial_t\gamma_+(\tau)(P_\ast(\tau,\yy,\xx,\uu))^2  d\tau\,\,d\yy\,d\xx \nonumber\\
&+\frac{L}{2}\int_\Omega\int_\Omega\,\rho^{\epsilon}(\yy,\xx)\left[g^-(r_*(t))-\frac{(\partial_{r^*}g^-(r_*(t)))^2}{2\overline{\mu}\gamma_-(t)}\right]\, d\yy d\xx\nonumber \\
&-\frac{1}{2}\int_\Omega\int_\Omega\int_0^t\,|\yy-\xx|\rho^{\epsilon}(\yy,\xx) \overline{\mu}\gamma_-(\tau)\dot{P}_\ast(\tau,\yy,\xx,\uu)P^\ast(\tau,\yy,\xx,\uu)\,d\tau\,d\yy\,d\xx\nonumber\\
&+ \frac{1}{2}\int_\Omega\int_\Omega\int_0^t\,|\yy-\xx|\rho^{\epsilon}(\yy,\xx)\overline{\mu}\partial_t\gamma_-(\tau)\left(\frac{\partial_rg^-(r_\ast(\tau))}{\overline{\mu}\gamma_-(\tau)\sqrt{|\yy-\xx|/L}}P^\ast(\tau,\yy,\xx,\uu)\right)-\frac{1}{2}\overline{\mu}\partial_t\gamma_-(\tau)(P^\ast(\tau,\yy,\xx,\uu))^2  d\tau\,d\yy\,d\xx \nonumber\\.
\end{align}



We conclude that the power balance delivers the energy balance given by:

\noindent{\bf Energy balance for fracture with damage and plasticity}
\begin{align}\label{eq:processzoneenergybalance2}
{\mathcal{D}}^\epsilon(t) =\int_0^t\left(\int_\Omega\,\bb(\tau)\cdot\dot\uu(\tau)\,d\xx+{\mathcal{R}}^\epsilon(\tau)\, \right)d\tau-({\mathcal{K}}(t) + {\mathcal{E}}^\epsilon(t) - {\mathcal{K}}(0) - {\mathcal{E}}^\epsilon(0)),
\end{align}
where $\mathcal{D}^\epsilon(t)$ is given by \eqref{damageenergy}.

The energy of the  subset of the specimen  where all the bonds have failed in tension and where all the bonds have failed in compression can be identified.
Let  $\Gamma^+(t)$ denote the set of bonds in the specimen that have failed in tension at time $t$ and $\Gamma^-(t)$ denote the set of bonds in the specimen that have failed in compression at time $t$. Clearly $\gamma_+(t,\yy,\xx,\uu)=0$, $\partial_t\gamma_+(t,\yy,\xx,\uu)=0$ and $\partial_{r^*}g^+(r_t^F)=0$ for all for all bonds in $\Gamma^+(t)$ and from \eqref{damageenergy} it is evident that the total energy dissipated by all bonds that have failed at time $t$ is
\begin{align}
\label{failtension}
{\mathcal{F}}_{tension}(t)=\frac{1}{2}\iint_{\Gamma^+(t)}\,\rho^{\epsilon}(\yy,\xx){L}\left[g^+(r_t^F)\right]\, d\yy d\xx.
\end{align}
Clearly $\gamma_-(t,\yy,\xx,\uu)=0$, $\partial_t\gamma_-(t,\yy,\xx,\uu)=0$, and $\partial_{r^*}g^-(r_c^F)=0$ for all for all bonds in $\Gamma^-(t)$ and from \eqref{damageenergy} it is evident that the total energy dissipated by all bonds that have failed in compression at time $t$ is
\begin{align}
\label{failcompression}
&{\mathcal{F}}_{compression}(t)=\frac{1}{2}\iint_{\Gamma^-(t)}\,\rho^{\epsilon}(\yy,\xx){L}\left[g^-(r_c^F)\right]\, d\yy d\xx.
\end{align}
It is clear that ${\mathcal{F}}_{tension}(t)$ and ${\mathcal{F}}_{compression}(t)$ are non-decreasing in time and will be calculated for tension and compression cracks in the next section.

\section{Calibration of nonlocal model using measured properties of the material} 
\label{sec.Straight Cracks}

In this section, we calibrate the model using quantities obtained directly from the dynamics and equating them to the material properties of the specimen. The elastic moduli and critical energy release rate follow from rigorous mathematical analysis of the IVBP on taking the $\epsilon=0$ limit of the peridynamic energies as shown in  \cite{lipton2016cohesive}, \cite{lipton2024energy}, \cite{mesoscopic}.

The elastic moduli are obtained here  formally by Taylor series and the energy release rate is recovered rigorously using geometric measure theory arguments of \cite{lipton2024energy}. 



\subsection{Elastic properties}

The Lam\'e constants for undamaged material are used to calibrate $\overline{\mu}$. This corresponds to the collapsed peridynamic tensor and energy release rate in the $\epsilon=0$ limit of the Peridynamic energy discovered in \cite{lipton2014dynamic}, \cite{lipton2016cohesive}. Here we use Taylor series implicitly and follow Section 6.5.2  of \cite{lipton2016cohesive}. 
Inside undamaged material containing a neighborhood $\mathcal{H}_\epsilon({\bf x})$ the elastic energy density is given by
\begin{align}\label{elastictundamaged}
\mathcal{W}^\epsilon(t,{\bf x})=\frac{1}{2}\int_{\Omega} \rho^\epsilon(\yy,\xx)|\yy-\xx|\overline{\mu} \frac{S^2(\yy,\xx,\uu(t))}{2} \,d\yy.
\end{align}
For a neighborhood ${\mathcal{H}}_\epsilon(\xx)$ chosen in undamaged material we assume the displacement is elastic and the horizon is small enough so that to first order the displacement is linear everywhere in ${\mathcal{H}}_\epsilon(\xx)$.
For this case $\uu=M{\bf x}$ where $M$  is a constant matrix, then ${S(\yy,\xx,\uu(t))}=F\ee\cdot \ee$, where $F=(M+M^T)/2$. Changing variables $\yy=\xx+\epsilon\boldsymbol{\zeta}$, with $|\boldsymbol{\zeta}|<1$, the elastic energy density over a neighborhood of $\xx$ at  time $t$ is
 \begin{align}
\mathcal{W}^\epsilon(t,{\bf x})&=\frac{1}{4\omega_d}\int_{\mathcal{H}_1(0)}|\zeta| J(|\zeta|)\overline{\mu}(F\ee\cdot\ee)^2\,d\zeta,
\label{LEFMequality}
\end{align}
with $\boldsymbol{\zeta}$ a vector inside the unit ball $\mathcal{H}_1(0)$, $d\zeta$ the volume element of the unit sphere, and $\ee$ a vector on the surface of the unit sphere. Observe that $(F\ee\cdot \ee)^2=\sum_{ijkl}F_{ij}F_{kl}e_ie_je_ke_l$ so
\begin{eqnarray}
\mathcal{W}^\epsilon(t,{\bf x})=\frac{1}{2}\sum_{ijkl}\mathbb{C}_{ijkl}F_{ij}F_{kl}
\label{leading order}
\end{eqnarray}
where
\begin{eqnarray}
\mathbb{C}_{ijkl}=\frac{\overline{\mu}}{2\omega_d}\int_{\mathcal{H}_1(0)}|\xi|J(|\xi|)\,e_i e_j e_k e_l\,d\xi=\frac{\overline{\mu}}{2\omega_d}\int_0^1\int_{\mathcal{H}_1(0)}|\xi|J(|\xi|)\,d|\xi|\int_{S^{d-1}}\,\Gamma_{ijkl}(\ee)\,de,
\label{elasticpart4}
\end{eqnarray}
with $de$ being the surface area element of the unit sphere $S^{d-1}$ in $d$ dimensions and $\ee$ a unit vector on $S^{d-1}$, $Q$ any rotation matrix representing an element of the $d$ dimensional rotation group $SO(d)$ and
\begin{equation}
\Gamma_{ijkl}(\ee)=e_i e_j e_k e_l \,\,\,\,\hbox{with} \quad \Gamma_{ijkl}(Q\ee)=Q_{im}e_m Q_{jn}e_n Q_{ko}e_o Q_{lp}e_p
\label{elasticpart5}
\end{equation}
To see that $\mathbb{C}_{ijkl}$ is an isotropic fourth order tensor we appeal to \eqref{elasticpart5} and \eqref{elasticpart4} to observe that $\mathbb{C}$ possesses the symmetry given by
\begin{align}
\label{elasticpart6}
\mathbb{C}_{ijkl}=Q_{im} Q_{jn} Q_{ko} Q_{lp}\mathbb{C}_{mnop},
\end{align}
for all rotations in $SO(d)$ to conclude that $\mathbb{C}$ is isotropic, so
\begin{eqnarray}
\mathbb{{C}}_{ijkl}=a (\delta_{ik} \delta_{jl} +\delta_{il}\delta_{kj})+ b\delta_{ij}\delta_{kl}.
\label{elasticpart7}
\end{eqnarray}
We evaluate $a$ by contracting both sides of \eqref{elasticpart4} with a trace free matrix and $b$ by
contracting both sides with the $d\times d$ identity and calculation delivers $a=b=\frac{\overline{\mu}}{8}\int_{0}^1r^2J(r)dr$ for $d=2$ and $a=b=\frac{\overline{\mu}}{10} \int_{0}^1r^3J(r)dr$ for $d=3$.
On the other hand, the potential elastic energy per unit volume of a material characterized by shear moduli $G$ and Lam\'e constant $\lambda$ is given by
\begin{align}
U&=\frac{1}{2}\sum_{ijkl}\mathbb{\tilde{C}}_{ijkl}F_{ij}F_{kl}
\label{LEFMequality13}
\end{align}
where
\begin{eqnarray}
\mathbb{\tilde{C}}_{ijkl}=G (\delta_{ik} \delta_{jl} +\delta_{il}\delta_{kj})+ \lambda\delta_{ij}\delta_{kl}
\label{elasticpart7}
\end{eqnarray}
Setting $U=\mathcal{W}^\epsilon(t,{\bf x})$, we get $\mathbb{\tilde{C}}=\mathbb{C}$ and we arrive at the calibration for determining $\overline{\mu}$ given by
\begin{eqnarray}\label{lambdamu}
G=\lambda=\frac{\overline{\mu}}{8}\int_{0}^1r^2J(r)dr, \hbox{ $d=2$}&\hbox{ and }\\
G=\lambda=\frac{\overline{\mu}}{10} \int_{0}^1r^3J(r)dr, \hbox{ $d=3$}.\label{calibrate1}
\end{eqnarray}
With Possion ratio $\nu=1/3$ for $d=2$ and $\nu=1/4$ for $d=3$. These formulas are consistent with those obtained from $\Gamma$-convergence of peridynamic energies obtained in \cite{lipton2016cohesive}.
For more general bond models the elastic  moduli are deduced by distinct but related means in, \cite{HuLi1} and \cite{HuLi2}.

 \subsection{Strength}
The domain of elastic behavior is calibrated to that of the material.  Here, we use the tensile stress of a material corresponding to the onset of nonlinear behavior $\sigma^C$. 
The critical stain in tension $r_t^C$ is determined by
\begin{align}\label{strength}
\sigma_t^C=g'(r_t^C)/\sqrt{|\yy-\xx|/L}.
\end{align}
The critical stain in compression $r_c^C$ is determined by
\begin{align}\label{strengthcompress}
\sigma_c^C=g'(r_c^C)\sqrt{|\yy-\xx|/L}..
\end{align}
The stain in tension $r_t^Y$ where the bond stiffness departs from linear behavior and exhibits plastic stiffening is determined by
\begin{align}\label{yieldstrengthstrain}
\sigma_t^Y=\overline{\mu} S_t^Y.
\end{align}
The stain in compression $S_c^Y$ where the bond stiffness departs from linear behavior and exhibits plastic stiffening is determined by
\begin{align}\label{yieldstrengthcompress}
\sigma_c^Y=\overline{\mu} S_c^Y.
\end{align}

\subsection{Critical energy release rate and determination of characteristic dimension }
\label{criticalenergyrelease}

The measured value of the critical energy release rate of the material is used to calibrate the
nonlocal constitutive parameters. 
First suppose the critical energy release rate for a mode I crack is given and the value is denoted by $G_c^+$. We then use \eqref{failtension} and calculate the dissipation energy of a flat crack. The mode 1 crack is denoted by $R_t$ and is a line segment for $d=2$ and a piece of flat surface in $d=3$. The domains $\Gamma^+(t)$ and $\Gamma^-(t)$ for a line crack is illustrated in Figure \ref{P}. The length of a line segment is written as ${\mathcal{H}}^1(R_t)$ and the area of a flat piece of surface is ${\mathcal{H}}^2(R_t)$.
Explicit computation as in \cite{lipton2024energy} using Crofton's formula gives
\begin{align}
\label{failtensionsimple}
{\mathcal{F}}_{tension}(t)=&\frac{1}{2}\iint_{\Gamma^+(t)}\,\rho^{\epsilon}(\yy,\xx){L}\left[g^+(r_t^F)\right]\, d\yy d\xx\nonumber\\
&=\mathcal{G}_c^+\times\mathcal{H}^{d-1}(R_t).
\end{align}
Its clear that this is a fracture energy and the energy release rate
$\mathcal{G}^+_c$ is given in terms of the constants used in the nonlocal model and is
\begin{align}
\mathcal{G}_c^+=\frac{L g^+(r_t^F)}{2}\frac{2\omega_{d-1}}{\omega_d}\, \int_{0}^1 r^{d}J(r)dr, \hbox{  for $d=2,3$}
\label{epsilonfracttough}
\end{align}
where $\omega^{d-1}$ and $\omega^d$ are the areas of the unit ball in dimension $d-1$ and  $d$ respectively. To calibrate we set $G_c^+={\mathcal{G}}_c^+$.

Similarly we can consider compression and mode II sliding cracks with prescribed critical energy release rate $G_c^-$.
Explicit computation as before gives
\begin{align}
\label{failcompressionsimple}
{\mathcal{F}}_{compression}(t)=&\frac{1}{2}\iint_{\Gamma^-(t)}\,\rho^{\epsilon}(\yy,\xx){L}\left[g^-(r_c^F)\right]\, d\yy d\xx\nonumber\\
&=\mathcal{G}_c^-\times\mathcal{H}^{d-1}(R_t).
\end{align}
As before its clear that this is also a fracture energy and the energy release rate
$\mathcal{G}^-_c$ is given in terms of the constants used in the nonlocal model and is
\begin{align}
\mathcal{G}_c^-=\frac{L g^-(r_c^F)}{2}\frac{2\omega_{d-1}}{\omega_d}\, \int_{0}^1 r^{d}J(r)dr, \hbox{  for $d=2,3$}.
\label{epsilonfracttough2}
\end{align}
To calibrate we set $G_c^-={\mathcal{G}}_c^-$. 

In this treatment we illustrate the ideas numerically for materials that fail in tension and are linear elastic in compression. In all simulations we apply the bilinear model and choose influence function $J(r)$ given by $1$ for $0<r<1$ and zero for $1\leq r$ and for dynamic simulations $L$  is determined by setting $G_c^+={\mathcal G}_c^+$ and we get
\begin{align}
    \label{lenthscale}
    L=\frac{3\pi G_c^+}{r_t^F\sigma^C},\,\,\,r_t^F\geq r_t^C=\frac{\sigma^C}{9E}.
\end{align}
An analogous equation holds for compression.
\subsection{Unloading ratio}
\label{unloadingratio}
The unloading ratio provides a coarse scale means for comparing how different microscale physics effects coarse scale behavior. For quasi-brittle fracture it is a measure of the ratio of irreversible strain due to frictional sliding over microscopic cracks to the irreversible strain consistent with the combined effects of frictional sliding and  degradation of elastic properties through creation of micro-voids and opening micro-cracks, \cite{ZhyShaoKobdo}. One can compute this number from available experimental results. 
Here the crack mouth opening displacement(CMOD) data for $3$ point bending tests versus  force history is used. It is obtained from boundary loads centered over the single pre-crack, see Figure \ref{figure:chenLiu-modeI}. As an example in the next section we pull the unloading ratio from the cyclic load-CMOD curve for a specific batch of concrete made and tested in \cite{chen2023fracture}, see Figure \ref{chen-liu force vs cmod}. Here an unloading curve highlighted in Figure \ref{fig:beta calibration} is selected. The $x$ axis intercept  $x_1$ is the plastic strain with elastic softening for the batch. We then choose the common point $P$ on the loading curve and construct a line through this point of slope equal to $\overline{\mu}$ and calculate the $x$ axis intercept $x_2$ of this line. The unloading ratio for the batch is $\beta_+=x_1/x_2$. An analogous calibration using compression loading determines the unloading ratio under compression $\beta_-$.

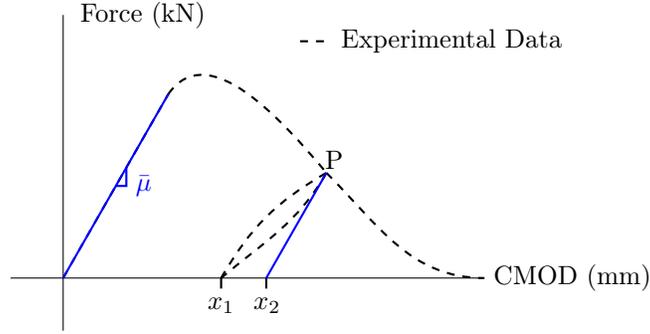
\begin{figure}
    \centering
    \begin{tikzpicture}[scale=0.7]

\draw[thin] (-1,0) -- (8,0);   
\draw[thin] (0,-1) -- (0,5);    
\draw[dashed, thick] (4.5,4.5) -- (5.1,4.5);
\node[right] at (5.1,4.5) {Experimental Data};

\draw[dashed, thick] (0,0)--(2,3.5);
\draw [dashed,thick] (2,3.5) to [out=55,in=130] (5,2); 

\draw [dashed,thick] (5,2) to [out=-120,in=40] (3,0); 

\draw [dashed,thick] (3,0) to [out=60,in=210] (5,2); 

\draw [dashed,thick] (5,2) to [out=-45,in=180] (8,0);

\draw[blue, thick] (0,0)--(2,3.5);

\draw[blue, thick] (1,1.75)--(1.2,1.75)--(1.2,2.1); 

\node[above] at (5.15,1.9) {P};

\node[right] at (1.2,1.75) {\color{blue}{$\bar{\mu}$}}; 
\draw[blue, thick] (5,2) to (3.857142,0);
\draw[black, thick] (3.857142,0) to (3.857142,-0.2);

\node[below] at (3.875142,-0.2) {\color{black}{$x_2$}}; 

\draw[black, thick] (3,0) to (3.,-0.2);

\node[below] at (3,-0.2) {\color{black}{$x_1$}}; 

\node[right] at (0.15,5) {\text{Force (kN)}};
\node[right] at (8,0) {\text{CMOD (mm)}};

\end{tikzpicture}
    \caption{Calibration of $\beta_\pm$ based on experimental data. $x_1$ is the $x$ axis intercept of a chosen unloading curve as measured by experiment and $x_2$ is the point of unloading that would occur without any loss in elasticity. The ratio of these values is the unloading ratio, $\beta_\pm = x_1/x_2$}.
    \label{fig:beta calibration}
\end{figure}

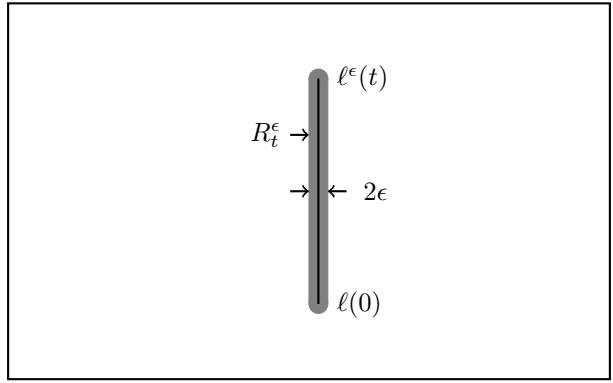
\begin{figure} 
\centering
\begin{tikzpicture}[xscale=0.50,yscale=0.50]


\node[left] at (-0.5,1.5) {$R_t^\epsilon$};

\draw [->,thick] (-0.5,1.5) -- (-0.0,1.5);




\draw [thick] (-8,-5) rectangle (8,5);









\node [right] at (0.5,3.0) { $\ell^\epsilon(t)$};

\draw [fill=gray, gray] (0.5,-3.0) rectangle (0.0,3.0);
\draw[fill=gray, gray] (0.25,3.0) circle (0.25);


\node [right] at (0.5,-3.0) { $\ell(0)$};
\draw[fill=gray, gray] (0.25,-3.0) circle (0.25);
\draw [thick] (0.25,-3.0) -- (0.25,3.0);
\draw [->,thick] (-0.5,0.0) -- (-0.0,0.0);
\draw [<-,thick] (0.5,0.0) -- (1.0,0.0);
\node [right] at (1.2,0.0) {$2\epsilon$};
\end{tikzpicture} 
\caption{{  Flat cracks and bond alignment. 
The crack $R_t^\epsilon=\{x_1=0;\,\ell(0)<x_2<\ell(t)\}$ and zone $\Gamma^\pm(t)$ where at least one bond is experiencing plastic strain and $\gamma_\pm(t,\yy,\xx,\uu)<1$ is given by the gray shaded region. We call this alignment/localization because bonds experiencing plastic strain and $\gamma_\pm(t,\yy,\xx,\uu)<1$ cross the flat crack set $R_t$. The energy of all the broken bonds are integrated up in the calculation of  the failure energy ${\mathcal{F}}(t)$} associated with the crack.}
 \label{P}
\end{figure}


\section{Results}
\label{sec.numerics}
This section provides the details on how the model is implemented in computational simulations. A mesh convergence study is carried out to ensure numerical accuracy. Following this study, the numerical experiments are carried out and compared with true experimental data across four benchmark problems: (i) monotonic and cyclic Mode I fracture, (ii) cyclic mixed-mode fracture, (iii) mixed-mode fracture induced by a corner singularity, (iv) the size effect in concrete beams, and (v) comparison with other dynamic phase field fracture simulations. 
 All simulations use a simple bilinear strength envelope to describe the constitutive law throughout this section, see \eqref{bilinear}. Eq. \eqref{gamma} provides the explicit form of the two-point phase field used within this study. The details of the calibration of the material constants are provided in Section \ref{sec.Straight Cracks}. The rational behind using a simple strength envelope is to demonstrate that the method gives solid agreement with experimental results and other phase field methods for the simplest nonlinear constitutive law between two point force and strain. 

The quasi-static results were obtained using the dynamic relaxation method \cite{underwood1983}. A damping term is added to the equation of motion given in \eqref{eq: linearmomentumbal2}, and the central difference time integration algorithm is used to obtain the displacement field satisfying the static equilibrium at each time step. However, the time step corresponds to the loading step rather than the physical time step used in transient analyses. Since this technique is an explicit iterative algorithm, the critical time step should be calculated accordingly. In this study, the critical time step is calculated as described by \cite{sillingaskari2005}. The numerical algorithm for the utilized dynamic relaxation method is provided in the supplementary materials of \cite{coskundamerchelilipton}.

This study focuses on the quasi-brittle failure in which tension softening and plastic deformation occur. For numerical simulation we choose a bilinear failure envelope given by \eqref{bilinear}. Hence, the bond force densities can be calculated using \eqref{Eqn.const}. Using \eqref{strains}, and \eqref{lambdamu}, with an influence function of $J(r)=1$ for two-dimensional domains, the calibration of the constant $\overline{\mu}$ results in $\overline{\mu} = 9E$.
where $E$ is the Young's Modulus of the material. Then, using \eqref{strength}, and \eqref{bilinear} \eqref{strains}, 
the critical bond strain is
\begin{equation} \label{sc}
    S^C = \sqrt{\frac{L}{|\yy-\xx|}} \frac{\sigma^C}{9E},
\end{equation}
\noindent
where $\sigma^C$ is the tensile strength of the material, and $L$ is the characteristic length of the domain. For static quasi-brittle failure, the characteristic length is the smallest dimension of the domain. The failure bond strain is calibrated using \eqref{epsilonfracttough}, \eqref{bilinear}, and is

\begin{equation}\label{sf}
    S^F = \frac{3 \pi \; {G}_c}{\sqrt{L |\yy-\xx|} \; \sigma^C}
\end{equation}

\noindent where ${G}_c$ is the critical energy release rate of the material, and \eqref{sf} is consistent with  \eqref{lenthscale}.  
Only uniform meshes and 2D domains are considered here. The displacement boundary conditions are applied at least to the horizon size of the layer nodes. The damping coefficient is selected as $2.0\times 10^7$ kg$/$m$^3$s unless otherwise stated. Finally, the horizon size is selected as 3.015 times of the uniform grid size. As \cite{trageser2020bond} stated earlier, Poisson's ratio is limited inherently due to the two-parameter formulation of the bond-based peridynamics. Therefore, Poisson's ratio is used as 0.33 throughout this study.
The data is extracted from the experimental literature using WebPlotDigitizer described in \cite{WebPlotDigitizer}.

\subsection{Mesh Convergence Study}
\label{sec.meshConvergence}
The mesh convergence study was conducted using a three point bending test (TBP). The material properties are given in Table \ref{table:chenLiu-material} and the dimensions of the specimen are given in Figure \ref{figure:chenLiu-modeI}. Only a uniform grid spacing was used to conduct the study so the analysis of things such as non-uniform, and mesh orientation are not addressed here.
\begin{figure}[H] 
	\centering
	\includegraphics[width=0.65\textwidth]{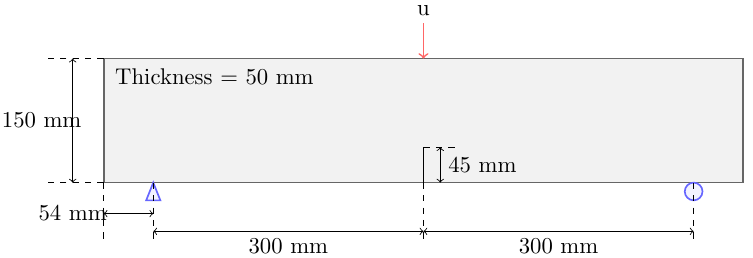}
	\caption{Dimensions and boundary conditions of the mode-I fracture test with a high strength concrete beam.}
	\label{figure:chenLiu-modeI}
\end{figure}
\begin{table}[H] 
	\small
	\centering
	\caption{Material constants used in the mode-I fracture test of the high-strength concrete beam.}
	\begin{tabular}{cccc}
		\hline \hline
		&Symbol & Values &Units \\
		\hline
		&$E$ & $33.89$ & GPa\\
		&$\sigma^C$ &  $3.39$ &  MPa\\
		&$\mathcal{G}_c$ &  $116.15$ &  N/m\\
		\hline\hline
	\end{tabular}
	\label{table:chenLiu-material}
\end{table}

Two convergence types were analyzed for this study: $\epsilon$-convergence and $m$-convergence. $\epsilon$ denotes the horizon and $m$ denotes the ratio of the horizon to grid spacing. Hence, we have $\epsilon = hm$ where h is the size of the uniform grid. $\epsilon$-convergence considers a decreasing horizon with a fixed $m$ value. 
$m$-convergence considers an fixed horizon and increasing $m$ value. When $m\to\infty$, the numerical solution is a discrete approximation of field theory. 
The simulated data can only be compared with the experimental data shown in Figure \ref{chenLiu-monolithic}. 

For the $m-$convergence, the horizon was fixed at $4 ,6,$ and $8$ mm and the m value is increase from $2$ mm to $8$ mm. Additionally, A more in-depth analysis was conducted for a horizon fixed at 6 mm with five $m$ values selected from $m=2$ to $m=8$. Figure \ref{m-convergence test} shows the reaction force vs crack mouth opening displacement (CMOD) for each setup. From the results, the difference between $m = 4$ and $m = 8$ is minimal, and virtually unnoticeable when going from $m=6$ to $m=8$ indicating that increasing $m$ beyond $8$ is unlikely to provide additional insight and we can reasonably conclude that the proposed model converges as $m\to\infty$.

\begin{figure}[H]
     
  \centering

    \begin{subfigure}[t]{0.49\linewidth}
    \includegraphics[width=\textwidth]{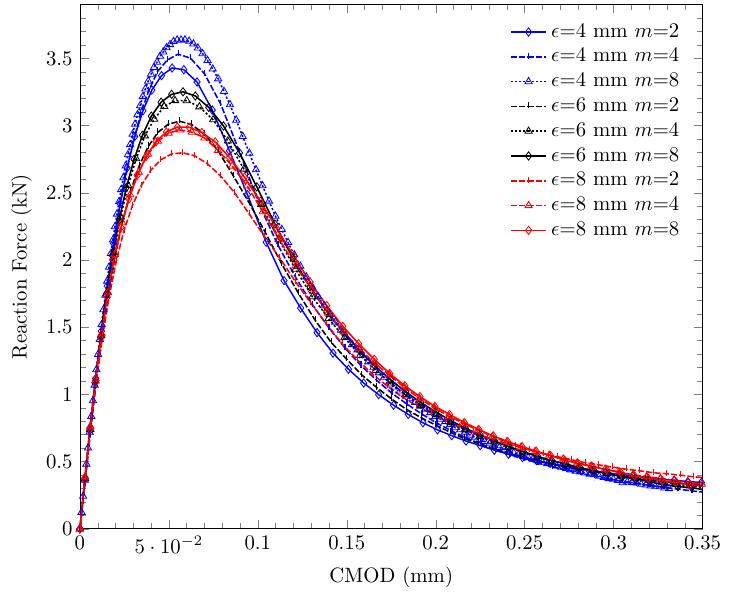}
        \caption{}
    \end{subfigure}
	\hfill
    \begin{subfigure}[t]{0.49\linewidth}
    \includegraphics[width=\textwidth]{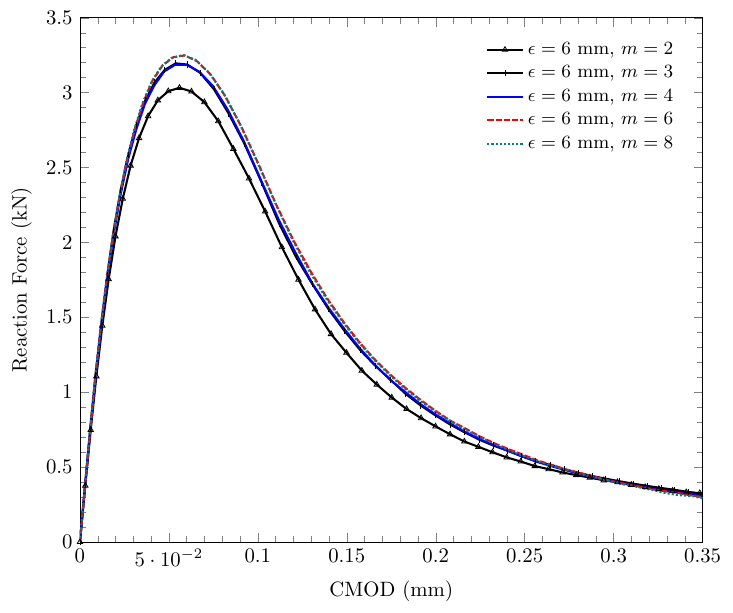}
        \centering
        \caption{}
    \end{subfigure}

   
    \caption{$m$-convergence test for $\epsilon=4$, 6, and 8}
    \label{m-convergence test}
\end{figure}

The $\epsilon-$convergence test followed a similar setup. The $m$ value was fixed at 2,4, and 8 and the horizon was decreased from 8 to 4 mm. In addition to this, a more in depth analysis was done for a fixed $m$ value of 4  with the horizon decreasing from 10 mm to 4 mm. The results  of these can be seen in Figure \ref{delta-convergence test}. When compared to the experimental results of \cite{chen2023fracture}, Our simulations recorded a maximum load of 2.8 kN for $\epsilon=10$ mm and $m=4$, a maximum load of 3.5 kN for $\epsilon=4$ mm and $m=4$, and the experimental data delivered a maximum load of 3.4 kN.

\begin{figure}[H]
  \centering

    \begin{subfigure}[t]{0.49\linewidth}
        \includegraphics[width=\textwidth]{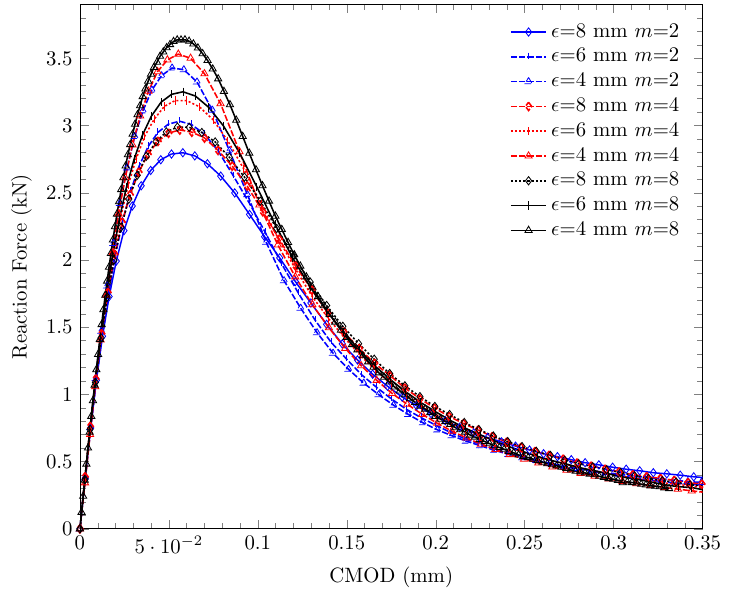}
        \caption{}
    \end{subfigure}
	\hfill
    \begin{subfigure}[t]{0.49\linewidth}
        \centering
        \includegraphics[width=\textwidth]{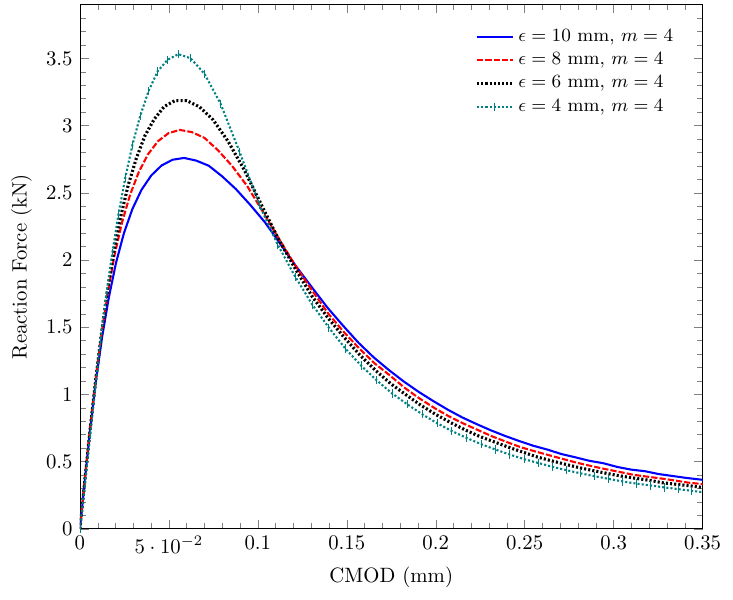}
        \caption{}
    \end{subfigure}

    \caption{$\epsilon$-convergence test for fixed $m=2$, $m=4$, and $m=8$.}
    \label{delta-convergence test}
\end{figure}

\subsection{Mode I fracture test for high-strength concrete under monolithic and cyclic loading}
\label{sec. hysterisiscrackgrowth}

The first numerical study was simulating mode I fracture under monolithic and cyclic loading. The material properties and the dimensional setup are shown in Figure \ref{figure:chenLiu-modeI} and Table \ref{table:chenLiu-material}. The material properties where pulled directly from \cite{chen2023fracture} and the results are directly compared with their experimental results. It should be noted that in the study conducted by \cite{chen2023fracture}, the loading was displacement controlled and the unloading was force controlled whereas for the numerical simulations, both loading and unloading are both displacement controlled. 

The horizon is selected to be 6 mm a with a uniform grid-spacing of 2 mm and a characteristic length of 150 mm. The stable time step was calculated to be $4.97\times10^{-7}$ s. The unloading ratio $\beta_+$ is calibrated from the experimental data given by the $5^{th}$ unloading-reloading cycle  for a centered pre crack.  From this calibration a $\beta_+$ value of $0.7$ was used. Figure \ref{chen-liu force vs cmod} presents the force vs CMOD for both monolithic loading and cyclic loading. The corresponding numerical simulations for both  monolithic and cyclic loads exhibit close agreement with the experimental results. 

Plastic strain concentration within the material can be assessed using the maximum value of plastic strain among all bonds within a horizon. Here we fix $\xx$ and measure $P^\ast(t,\yy,\xx,\uu)$ relative to the critical plastic strain $\beta S^F(\yy,\xx)$ among all bonds connected to $\xx$ within the horizon $\epsilon$ given by 
\begin{align}\label{Z}
   Z(t,\xx,\uu) = \max_{\yy\in H_\epsilon(\xx)}\left(\frac{r_p(t,\uu)}{\beta r^F} \right)=\max_{\yy\in H_\epsilon(\xx)}\left(\frac{P^*(t,\yy,\xx,\uu)}{\beta S^F(\yy,\xx)} \right).
\end{align}
The quantity $Z(t,\xx,\uu)$ is referred to as the strain concentration. The value of $Z(t,\xx,\uu)$ is greater than zero if there is a plastic strain between $\yy$ and $\xx$. It
takes the value one if at least one bond has failed.  
Figure \ref{fig:plastic-strain-contour} shows the level lines of $Z(t,\xx,\uu)$ localization and evolution of the plastic strain at four loading stages one before and three after the peak stress has been reached. $Z(t,\xx,\uu)$ and was plotted everywhere in the domain and is only nonzero around the crack and equal to $1$ adjacent to the crack.  The loading stages are shown in Figure \ref{chen-liu cyclic with laoding stages}.

\begin{figure}[H]
  \centering

    \begin{subfigure}[t]{0.49\linewidth}
        \includegraphics[width=\textwidth]{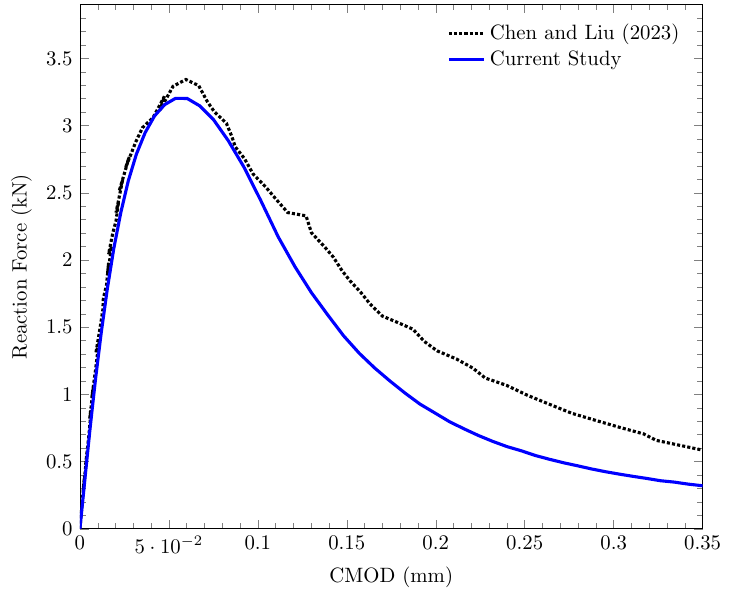}
        \caption{}
        \label{chenLiu-monolithic}
    \end{subfigure}
	\hfill
    \begin{subfigure}[t]{0.49\linewidth}
        \centering
        \includegraphics[width=\textwidth]{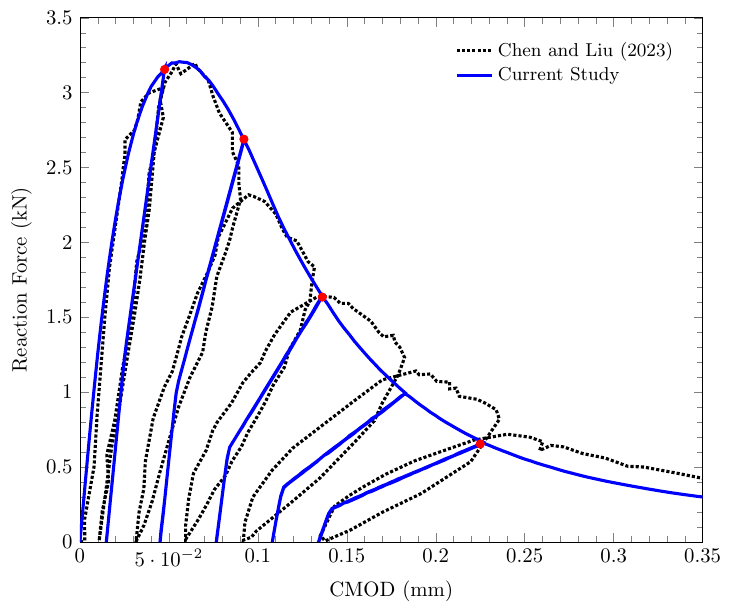}
        \caption{}
        \label{chen-liu cyclic with laoding stages}
    \end{subfigure}

    \caption{Comparison of Force vs CMOD for monolithic and cyclic loading  for the centered pre notch.}
    \label{chen-liu force vs cmod}
\end{figure}

\begin{figure}[!ht]
\centering

\includegraphics[width = \linewidth]{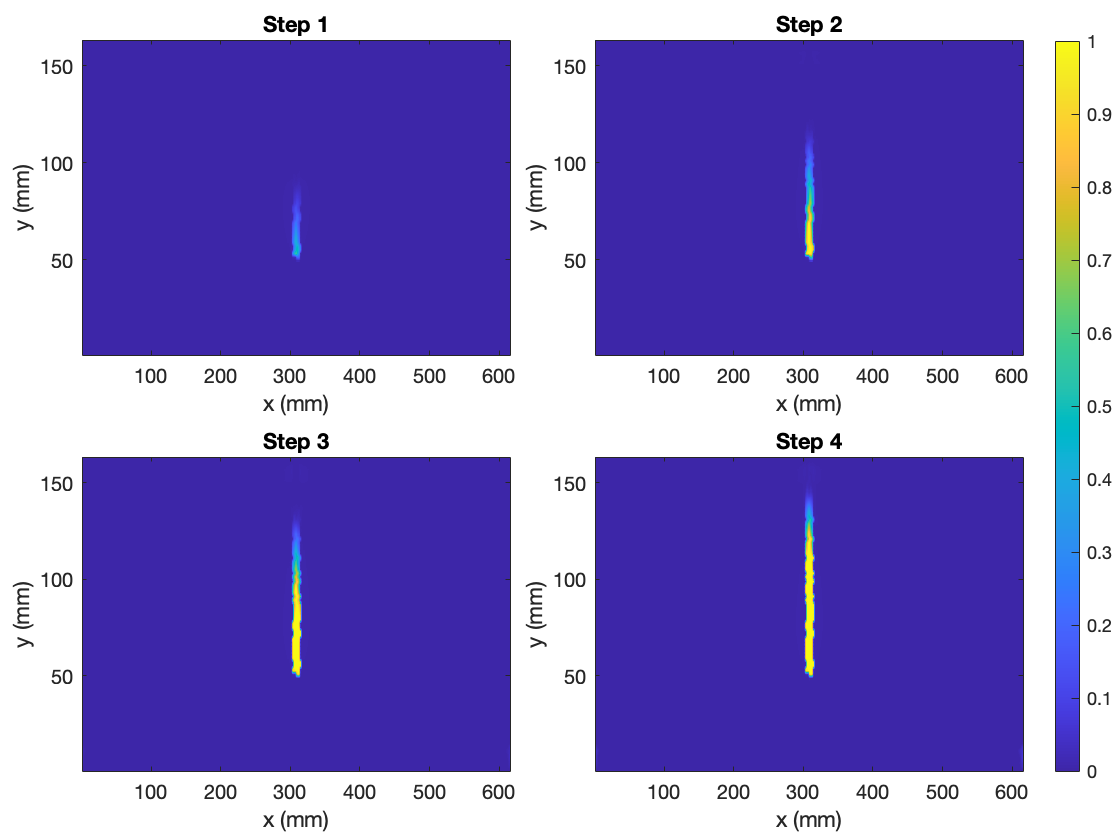}

\caption{Localization of the plastic strain inside specimen at four loading stages highlighted by the red dot in Figure \ref{chen-liu force vs cmod} b. Here we see localization of plastic strain around a growing quasi-brittle crack. The figures show that the crack is preceded by a process zone of finite size. Plastic strain is measured by $Z(t,\xx,\uu)$.  $Z(t.\xx,\uu)=0$ is rendered blue in the material with no damage and yellow when $Z(t,\xx,\uu)=1$ where al least one  bond has failed}
\label{fig:plastic-strain-contour}
\end{figure}

\subsection{Mixed-mode fracture of  concrete beam under cyclic loading}
\label{offcenter}

The next simulation is for mixed mode fracture of concrete beams where both shear and tensile stresses are present. This was carried out using a three point bending test with a prenotch that was offset at four different lengths. The dimensions of the beam and the material properties are selected to be consistent with \cite{jenq1988mixed} and are shown in Figure \ref{fig:geometry-mixedMode} and Table \ref{table:jenqShah-material}, respectively. However, the maximum tensile strength and the critical energy release rate are not specified by \cite{jenq1988mixed}, so these values are selected and remain unchanged throughout this section.
\begin{figure}[H]
	\centering
	\includegraphics[width=0.65\textwidth]{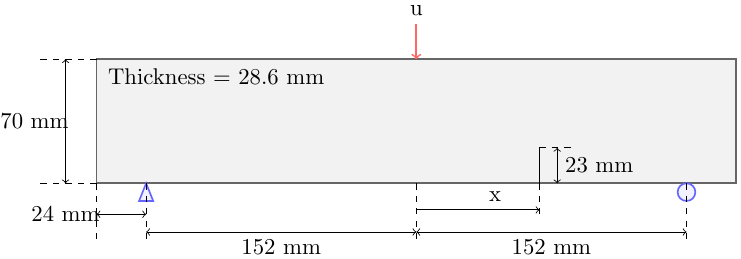}
	\caption{Dimensions and boundary conditions of the mixed-mode fracture tests on concrete beams.}
	\label{fig:geometry-mixedMode}
\end{figure}

\begin{table}[H]
	\small
	\centering
	\caption{Material constants used in the mixed-mode fracture tests of the concrete beams.}
	\begin{tabular}{cccc}
		\hline \hline
		&Symbol & Values &Units \\
		\hline
		&$E$ & $32.8$ & GPa\\
		&$\sigma^C$* &  $3.25$ &  MPa\\
		&$\mathcal{G}_c$* &  $115$ &  N/m\\
		\hline\hline
	\end{tabular}
	\vfill
	*Not provided in the experimental study by \cite{jenq1988mixed}.
	\label{table:jenqShah-material}
\end{table}

The horizon was selected to be 3 mm and the grid spacing was set at 1 mm. The time step was calculated to be $2.5\times 10^{-7}$ s and the characteristic length was set at 70 mm. $\beta_+$ for the \cite{jenq1988mixed} concrete specimens are calculated as in section \ref{unloadingratio} for a centered crack and found to be 0.722. We then set $\beta_+$ to this value for all simulations run in this section. The parameter $\gamma$ is defined as the ratio of the distance $x$ shown in Figure \ref{fig:geometry-mixedMode} to half the length of the specimen. This is used to describe offset distance of the prenotch. Figure \ref{jenq-shah force vs cmod} shows the force versus CMOD for four different values of $\gamma$.  We also calibrated $\beta_+$ for each different offset $\gamma$ and numerically computed the force versus CMOD for comparison. As expected the simulations become more accurate when this is done. 
\begin{figure}[H]
    \centering
    
    \begin{subfigure}{0.48\textwidth}
        \centering
       \includegraphics[width=\textwidth]{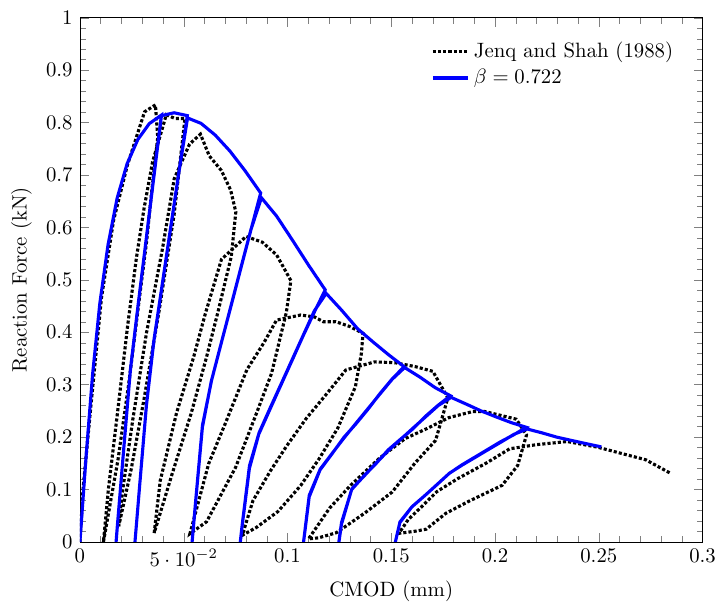}
        \caption{$x = 0$ mm, $\gamma=0$}
    \end{subfigure}%
    \hfill
    \begin{subfigure}{0.48\textwidth}
        \centering
        \includegraphics[width=\textwidth]{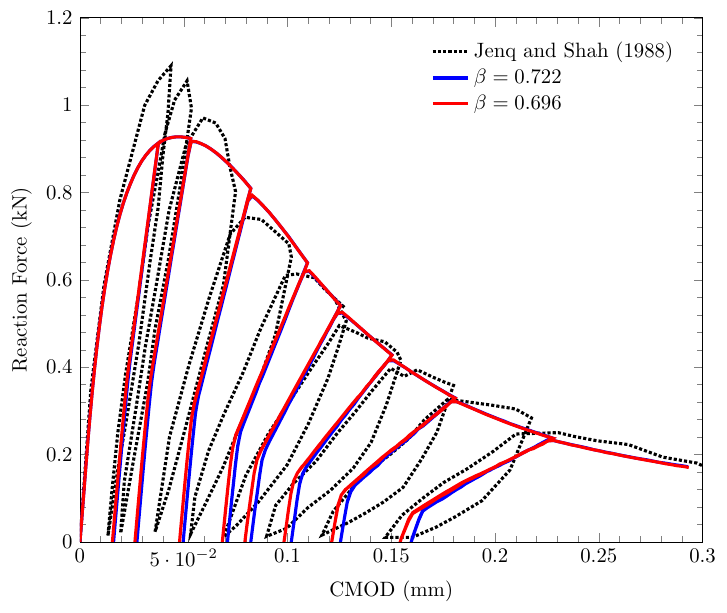}
        \caption{$x = 25$ mm, $\gamma=1/6$}
    \end{subfigure}

    \begin{subfigure}{0.48\textwidth}
        \centering
        \includegraphics[width=\textwidth]{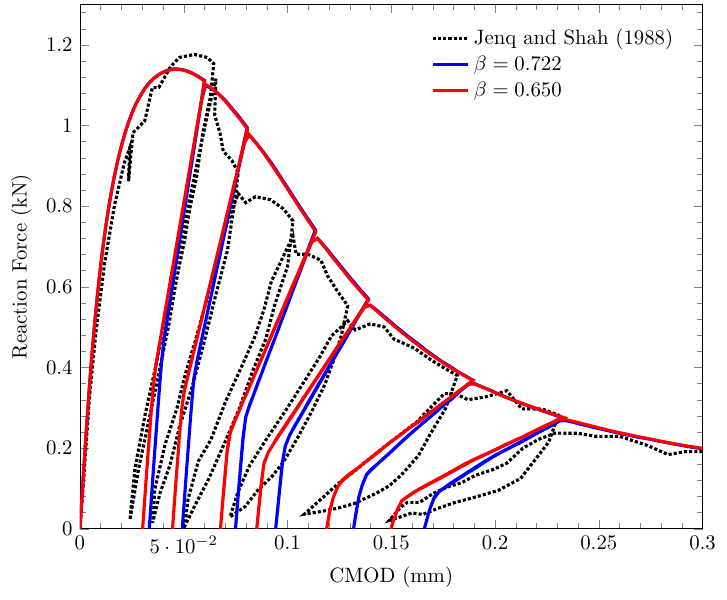}
        \caption{$x = 50$ mm, $\gamma=1/3$}
    \end{subfigure}%
    \hfill
    \begin{subfigure}{0.48\textwidth}
        \centering
       \includegraphics[width=\textwidth]{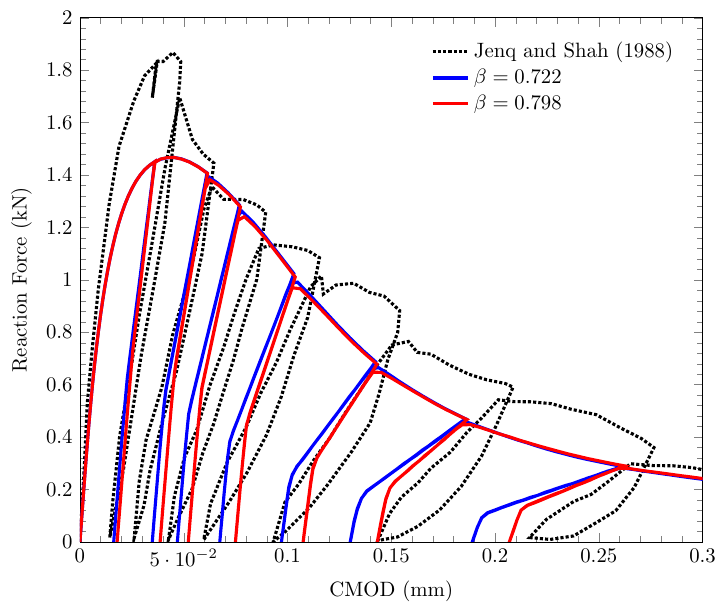}
        \caption{$x = 76$ mm, $\gamma=1/2$}
    \end{subfigure}

    \caption{Comparison of experimental and numerical results for various $\gamma$ values.}
    \label{jenq-shah force vs cmod}
\end{figure}
While there were four different sized specimens used in the original study, the experimental data for the Force vs CMOD curve is only given for a single study. Furthermore, the specific loading and unloading data is not given in the original study so the specific unloading points were selected to match the experimental data as closely as possible. Nevertheless, for the experimental data that is given, we see excellent agreement between the proposed model and the experimental data. 
Figure  \ref{fig:crack-paths} shows the crack path for each setup. The crack paths were pulled directly from the images of the damaged specimens shown in Figure 15 of \cite{jenq1988mixed}. The  displacement jumps in $x$ component of the computationally generated displacement field follows the direction of the experimental crack paths for the different offsets.

\begin{figure}[!h]
\centering

\begin{subfigure}{0.23\linewidth}
\centering
\includegraphics[scale = 0.5]{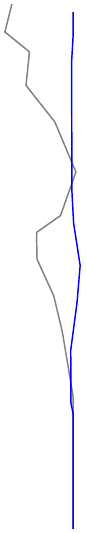}
\caption{$x=0$ mm}
\end{subfigure}
\hfill
\begin{subfigure}{0.23\linewidth}
\centering
\includegraphics[scale = 0.5]{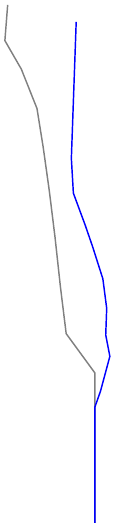}
\caption{$x=25$ mm}
\end{subfigure}
\hfill
\begin{subfigure}{0.23\linewidth}
\centering
\includegraphics[scale = 0.5]{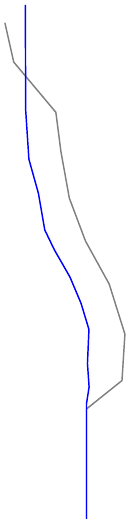}
\caption{$x=50$ mm}
\end{subfigure}
\hfill
\begin{subfigure}{0.23\linewidth}
\centering
\includegraphics[scale = 0.5]{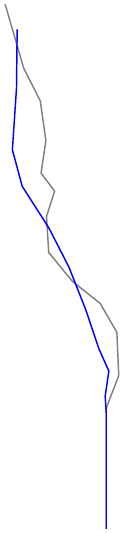}
\caption{$x=76$ mm}
\end{subfigure}

\caption{Comparison of experimental (gray) and numerical (blue) crack paths for different offsets.}
\label{fig:crack-paths}
\end{figure}

\subsection{Mixed mode fracture of L-shaped concrete specimen}
\label{mixedmode}
The next numerical example is conducted for an L-shaped concrete beam without a prenotch. The dimensions and loading setup are shown in Figure \ref{figure:lShapedDimensions} and the reference study was conducted by \cite{winkler2001traglastuntersuchungen}. The material properties are shown in Table \ref{table:jenqShah-material} and selected to be consistent with \cite{winkler2001traglastuntersuchungen} except for the elastic modulus which is selected to be 18 GPa  which better represents the initial slope.

\begin{table}[H] 
	\small
	\centering
	\caption{Material constants used in the mixed-mode fracture test of the concrete panel.}
	\begin{tabular}{cccc}
		\hline \hline
		&Symbol & Values &Units \\
		\hline
		&$E$* & $20$ & GPa\\
		&$\sigma^C$ &  $2.7$ &  MPa\\
		&$\mathcal{G}_c$ &  $90.0$ &  N/m\\
		\hline\hline
	\end{tabular}
	\vfill
	*Different than the experimental values provided by \cite{winkler2001traglastuntersuchungen}: E=25.85 GPa.
	\label{table:lShaped-material}
\end{table}
\begin{figure} 
	\centering
	\includegraphics[width=0.45\textwidth]{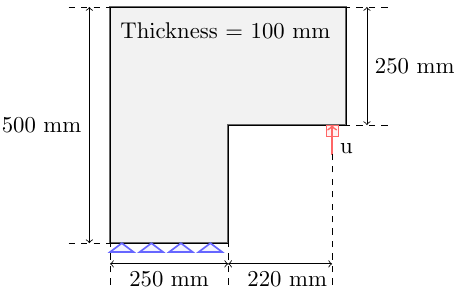}
	\caption{Dimensions and boundary conditions of the mixed mode fracture test with : L-shaped concrete panel.}
	\label{figure:lShapedDimensions}
\end{figure}

The boundary conditions fix the displacement in both the vertical and horizontal directions and the support is chosen to have a height equal to the horizon. The horizon is selected to be 7.875 mm with a gridsize of 2.5 mm. The local damping is set at $2.0\times 10^{7}$ kg/m$^3$ s, and the characteristic length is set to 250 mm. The critical timestep is calculated to be $9.04\times 10^{-7}$ s, and the specimen is loaded to a final displacement of 0.65 mm. Figure \ref{L-shaped force vs displacement} presents the reaction force vs displacement comparison for the numerical simulations and the experimental data. 

\begin{figure}[!ht]
\centering
\includegraphics[]{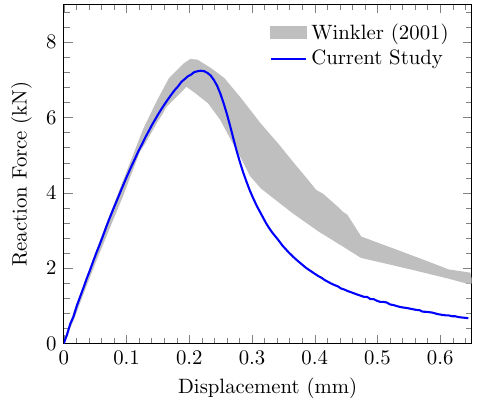}
\caption{Force vs Displacement for L-shaped Concrete Specimen}
\label{L-shaped force vs displacement}
\end{figure}

The ultimate load-carrying capacity and the subsequent decrease in load carrying capacity are predicted by the proposed model.
The maximum forces are essentially equal with both the experimental data and the numerical simulation, predicting a maximum force of roughly 7 kN. 
The post-peak softening behavior follows the general trend of the experimental data, though the numerical model predicts a slightly more rapid decrease in load capacity after 0.3 mm compared to the experimental range. On the other hand we are using a simple bilinear envelope for modeling the bond degredation and failure. The use of an exponential bond model with bonds that soften to zero at infinity will decrease the bond decay.

The crack path obtained from the numerical simulation is presented using a local damage variable. The damage variable can take any value between zero and one. Here zero indicates no softening of elastic properties and values greater than zero indicate softening and one indicates completely damaged points. Local damage at the material point $\xx$ at time \textit{t}, $D(\xx, t)$ can be calculated from the two point phase field as

\begin{equation} \label{eq:localDamage}
	D(\xx, t) = 1 - \frac{\int_{\HH_\epsilon(\xx)} \gamma_+(t,\yy,\xx,\uu) \,d\yy} {\int_{\HH_\epsilon(\xx)} \,d\yy},
\end{equation}
where $\gamma_+(t,\yy,\xx,\uu)$ is the two-point phase field described in Figure \ref{ConvexConcaveC}.
 Figure \ref{L-shaped crack path} shows the crack path predicted by the model and it compares within an acceptable accuracy to the experimental results of \cite{winkler2001traglastuntersuchungen}. 
We recover mode mixity along the crack by examining the jump discontinuities of the displacement across the crack. At the corner of the L-shaped domain it is a mode one jump progressing into a jump discontinuity given by a mixture of mode one and mode two (sliding) as one moves away from the corner along the crack path.
\begin{figure}[!ht]
\centering

\begin{subfigure}{0.44\linewidth}
    \centering
    \includegraphics[width=\linewidth]{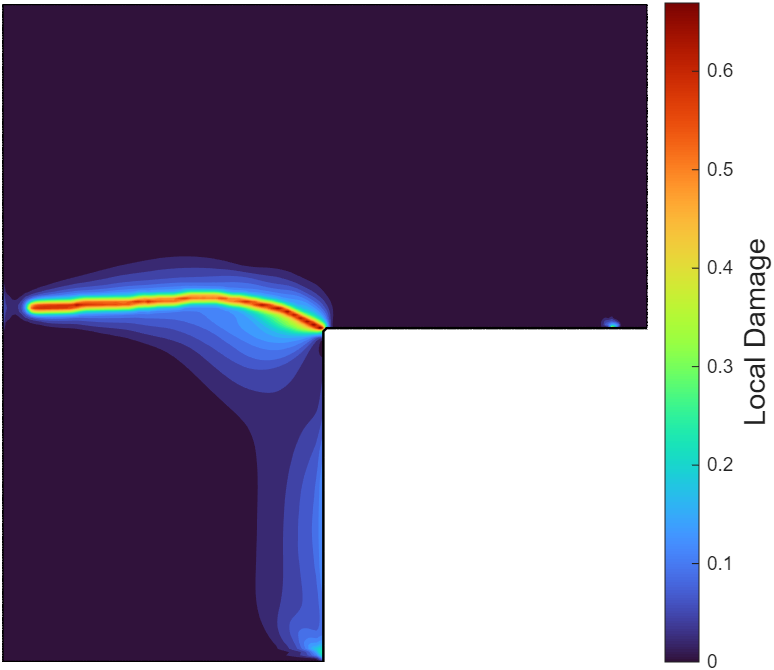}
    \caption{}
\end{subfigure}
\hfill
\begin{subfigure}{0.54\linewidth}
\centering
\includegraphics[width=\linewidth]{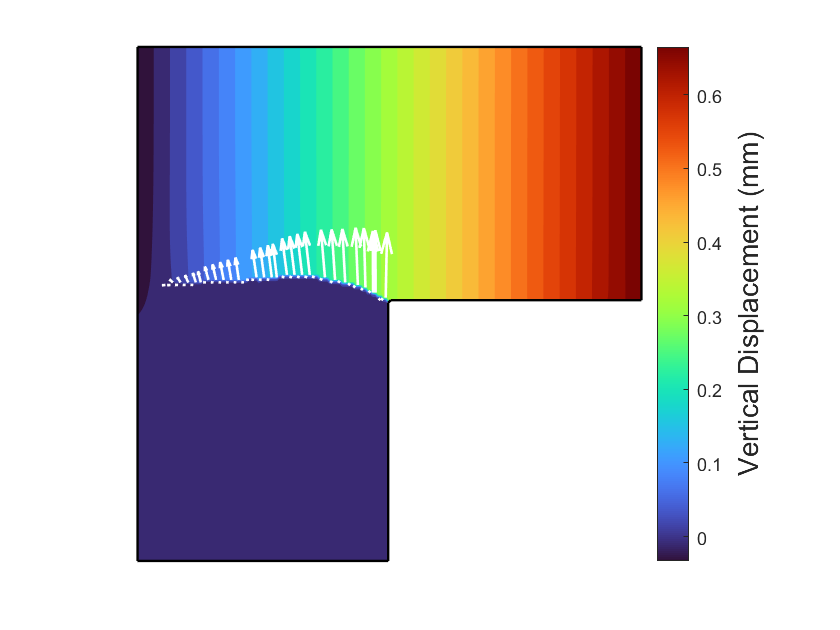}
\caption{}
\end{subfigure}
\caption{a) Crack path for L-shaped Concrete Specimen as presented by damage variable., b) Crack path presented by discontinuity set for displacement. The displacement field on either side of the crack represented by arrows. The displacement field jumps only on either side of the crack. Note that at the corner it is a mode one jump  going into a mixture of mode one and mode two (sliding) as one moves away from the corner along the crack path. }
\label{L-shaped crack path}
\end{figure}

\subsection{Size effect in concrete beams}
\label{sizeeffects}
The final numerical test examines the deterministic size effect in concrete beams. \cite{BazantPlanas1998} define the structural size-effect as the deviation of the actual load-carrying capacity of the structure, due to the change of its size, from the one predicted by any deterministic theory where the failure of the material is expressed in terms of stress and/or strain. Classical  plastic failure theories or theories based on maximum stress or strain predict that the stress at which a structure will begin to fail is independent of the size of that structure. This has been shown to be inaccurate through experimental testing. 

A mode I failure test was used to study the size effect and the numerical results were compared directly with the experimental results from \cite{garcia2012analysis}. Three different sized beams were used to carry out this study. Table \ref{table:sizeEffect-beamDim} gives the dimensions for each setup and Table \ref{table:sizeEffect-material} shows the material properties which were fixed for each beam and are consistent with the \cite{garcia2012analysis}.

\begin{table}[H] 
	\small
	\centering
	\caption{Dimensions of the beams tested for the mode-I failure test.}
	\begin{tabular}{cccccc}
		\hline \hline
		&Specimen &Depth (mm) &Span (mm) &Thickness (mm) &Prenotch Length (mm)\\
		\hline
		&Beam 1 &80  &200  &50 &20\\
		&Beam 2 &160 &400  &50 &40\\
		&Beam 3 &320 &800  &50 &80\\
		\hline\hline
	\end{tabular}
	\label{table:sizeEffect-beamDim}
\end{table}

\begin{table}[H] 
	\small
	\centering
	\caption{Material constants used in the mode-I fracture test of the concrete beams.}
	\begin{tabular}{cccc}
		\hline \hline
		&Symbol & Values &Units \\
		\hline
		&$E$ & $33.8$ & GPa\\
		&$\sigma^C$ &  $3.5$ &  MPa\\
		&$\mathcal{G}_c$ &  $80.0$ &  N/m\\
		\hline\hline
	\end{tabular}
	\label{table:sizeEffect-material}
\end{table}

The horizon is selected to be 6.03 mm with a uniform grid spacing of 2 mm. The local damping was set to $3.0\times 10^{7}$kg/m$^3$ s and the critical timestep was calculated to be $4.99\times 10^{-7}$ s. The characteristic length was set to 160 mm and was kept consistent across all three setups. Figure \ref{size effect:Force vs CMOD} shows the force vs CMOD curves for each beam size as well as the experimental results from and \cite{garcia2012analysis} and the results from another peridynamic model proposed by \cite{hobbs}. It should be noted that the experiments by \cite{garcia2012analysis} only include two experiments for beam 3 whereas beam 1 and 2 both have three experiments. 

\begin{figure}[!ht]
	\centering
\includegraphics[]{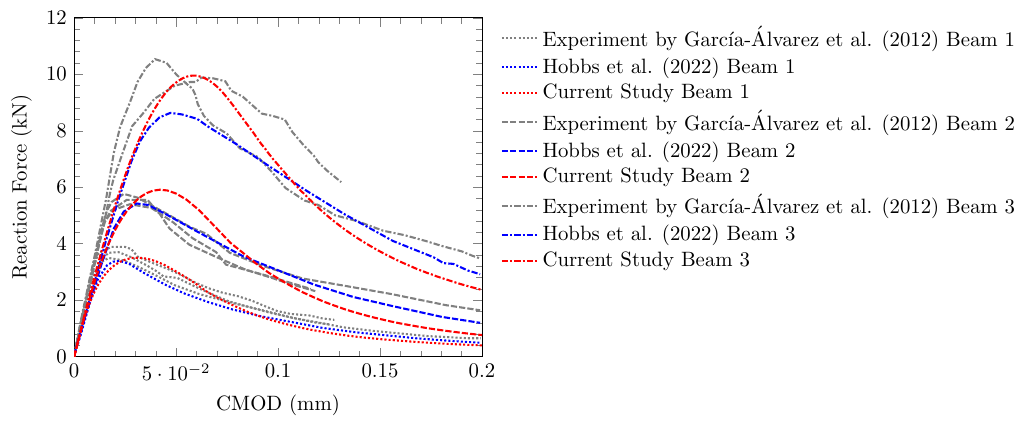}
\caption{Comparison of force vs CMOD curves for mode I fracture test}
\label{size effect:Force vs CMOD}
\end{figure}

The gradual decrease in reaction force for larger beam dimensions indicates that the model is capable of capturing the size effect. To further verify this, the ligament stress was calculated at the maximum load for each setup. The ligament stress can be calculated using 
\begin{equation} \label{eq:nominalStrength}
	\sigma_{\text{lig}} = 1.5 \frac{F_{ult} \; s}{b \; h^2}
\end{equation}
where $F_{ult}$ represents the maximum force carried, $s$ is the span of the beam, $b$ is the thickness of the beam, and $h$ is difference between the depth of the beam and the length of the prenotch. Once the ligament stress is calculated for each beam, it is plotted vs the beam depth in Figure \ref{figure:nominal stress size effect}. It is clear from the figure that the model clearly captures the size effects for each beam and is consistent with the experimental data. 

\begin{figure}[!ht]
	\centering
    \includegraphics[scale =1]{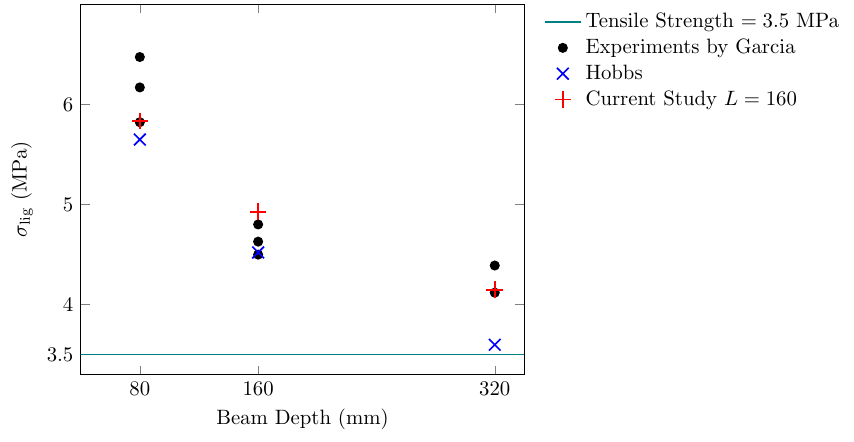}
\caption{Nominal stress values vs beam depth}	
\label{figure:nominal stress size effect}
\end{figure}

\subsection{Yield strength versus Griffith fracture criterion}
\label{strengthstability}
The final numerical study considered a single-edge notch tension test based on \cite{kamarei2026ninecircles}. The specimen geometry and dimensions were chosen to match those of the original study and are shown in Figure \ref{Figure:Yield strength vs Griffith dimensions}. The material properties used in the simulations are listed in Table \ref{table:chenLiu-material}.

\begin{figure}[H]
    \centering
\includegraphics[]{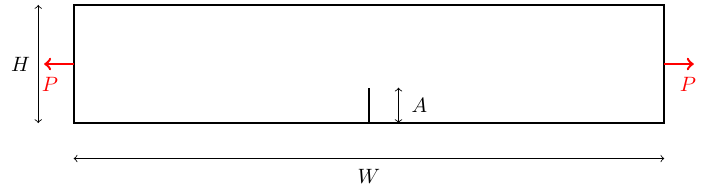}
\caption{Dimensions of single edged notch specimen where we pick $H=5$ mm, $W=25$ mm, $t=0.25$ mm, and $A=0.025,0.1, 0.5, 1, 1.5, 1.75, 2$ mm. Here $t$ denotes the thickness of the specimen}
\label{Figure:Yield strength vs Griffith dimensions}
\end{figure}

The objective of this study is to examine the dependence of the critical stress at fracture nucleation on the initial crack length. The strip is  clamped on left and right boundaries and pulled apart by an applied displacement and the flaw is a prenotch of length $A$. The resultant force is denoted by $P$. For sufficiently small flaws, fracture is expected to be governed primarily by the material yield strength. As the crack length increases, the critical stress of the specimen begins to decrease. According to LEFM, this decrease should approach the slope of the line given by
\begin{equation}
    S_{cr} = \sqrt{\frac{G_cE}{\pi A}}.
    \label{equation:LEFM bound}
\end{equation}

The gridspacing was set at $m = 0.15$ mm and the horizon was set at $\epsilon = 3.015m$. The characteristic length was set at $12.5$mm. Based on these parameters, the critical stable timestep was computed to be $2.57\times10^{-8}$ s. Figure \ref{yield strength vs griffith} shows the critical stress at fracture  predicted by the model and computed using $P/(Ht)$, as a function of the prenotch length A.

    \begin{figure}[H]
    \centering
    \includegraphics[]{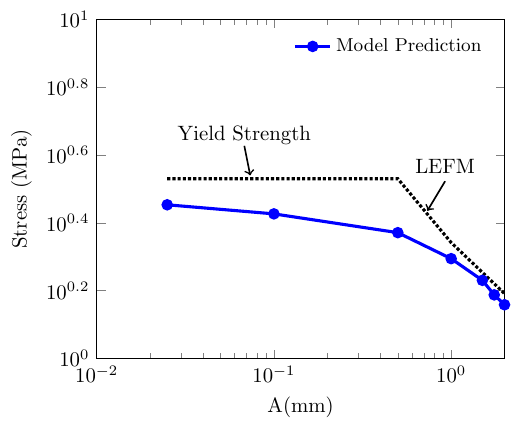}
        \caption{}
        \label{yield strength vs griffith}
    \end{figure}
For small prenotch lengths ((A = 0.025) mm and (A = 0.1) mm), the predicted critical stress remains close to the yield strength of the matieral. 
As the prenotch length increases, the critical stress values approach the slope of the line given by equation \ref{equation:LEFM bound}. This indicates that the model captures the expected transition from tensile-strength-controlled fracture for small flaws to LEFM-controlled behavior as the flaw size increases. This behavior of yield strength versus Griffith fracture criterion is consistent with the recent work of \cite{KumarFrancfortPalmes2}, \cite{kamarei2026ninecircles} and the size effect of \cite{BazantPlanas1998}.

\section{Conclusions}
\label{sec.conclusion}

The model is formulated as a mathematically well-posed initial value boundary value problem for the displacement inside a quasi-brittle material. The solution  delivers the displacement, damage field, irreversible strain and crack set history. The model is based on a blend of a nonlocal constitutive law and a two point  phase field. Displacement evolutions satisfy energy balance principles, with a positive energy dissipation rate that is consistent with thermodynamics. 
These properties are not postulated but are consequences of  the evolution equation.

In the numerical examples, the simplest possible constitutive rule is intentionally chosen to minimize the number of material and numerical constants. By utilizing the prescribed Young's modulus, tensile strength, fracture energy, and unloading ratio  the constants appearing in the numerical model can be determined. This delivers both qualitative and quantitative results that align closely with the experimental data. The numerical simulations demonstrate excellent agreement with the experimental results for several benchmark problems, including cyclic loading and mode-I fracture in concrete beams, cyclic loading and mixed-mode fracture in notched specimens, and  mixed mode fracture in L-shaped panels. In particular, the same set of material constants and constitutive model successfully capture the size-effect phenomenon across three different beam sizes without  any explicit rules for determining strength based on sample size. These results confirm the model's ability to replicate real-world material behavior with minimal computational complexity.
Most importantly in our framework  the length scale of non-locality ``$\epsilon$,'' {\em is not a material property and is not used in the calibration of strength, fracture energy, elastic or plastic properties}. The blended model is a  well-posed approximation theory under which the effects of damage and plasticity can localize onto surfaces. The length scale of localization is controlled by $\epsilon$.

In this paper we have worked in the simplest bond based context to demonstrate the usefulness of the method.  This basic work establishes the theoretical and computational validity of the blended approach for  physically meaningful problems and agrees well with experiment. It provides opportunity for {\bf prediction} of damage and plastic localization and the emergence of fracture. It also opens up an opportunity to extend these ideas to anisotropic media and composite materials using tensor valued phase fields and more general formulations of Newton's second law.

\section{Acknowledgment}
This material is based upon work supported by the U.S. Army Research Laboratory and the U.S. Army Research Office under Contract/Grant Number W911NF-19-1-0245 and W911NF-24- 2-0184.

\bigskip

\noindent {\bf Data availability} The code used to generate the results in this study is available at \\ https://github.com/SemsiCoskun/PDBlend.

\appendix

\section{Remarks on earlier work} 

We remark on the earlier work of \cite{She-Mo-Sh} motivated by the concrete damage models of \cite{AS} and others. That work provides algorithmic implementations of peridynamic models of cyclic loading with plastic effects. These models are close in spirit to the cyclic model presented here.  However a closer look shows an error in the constitutive models given by the Equations (15) and (16) of \cite{She-Mo-Sh}.  A straight forward application of algebra to the model shows these equations do not give admissible bond failure envelopes for two reasons:  1) there are bond force-strain pairs lying inside the failure envelope which do not correspond to unloading laws. 2) the slope of the unloading line is greater than that of the undamaged bond for a range of force-strain pairs in the post peak regime. These observations follow directly from the formula for the plastic strain given by $0.725\,s_{MAX}$ where $s_{MAX}$ is the strain associated with a stress on the backbone curve (loading curve), see Figure 3 or 4 of \cite{She-Mo-Sh}. To see the error set $S^C$ to be the critical strain when bonds begin to soften. When $s_{MAX}<S^C$ the plastic strain is zero. When $S^C<s_{MAX}$ and $s_{MAX}$ is just greater than the critical strain it is evident that the  plastic strain jumps from zero for $s_{MAX}<S^C$ to just greater than $0.725\,S^C$ thus the bond stiffness becomes greater than the undamaged bond, this implies bonds stiffen when they yield, which is a clear contradiction in view of Figures 3 and 4 of that paper. To fix this problem the bond plasticity can be correlated to material damage  proposed in the present article.



\bibliographystyle{elsarticle-harv} 
\bibliography{references}






\end{document}